\documentclass[prb,aps,epsf,showpacs,twocolumn]{revtex4}
\usepackage{times}
\usepackage{graphicx}
\usepackage{float}
\usepackage{latexsym,amsmath,amssymb,bm,euscript}
\usepackage{color}
\usepackage{subfigure}
\usepackage{epstopdf}
\usepackage[colorlinks=true,linkcolor=blue,citecolor=blue]{hyperref}
\usepackage{hyperref}

\begin{document}

\title{Global Phase Diagram of Competing Ordered and Quantum Spin
  Liquid Phases on the Kagom\'e Lattice}
\author{Shou-Shu Gong$^{1}$, Wei Zhu$^{1}$, Leon Balents$^{2}$, and D. N. Sheng$^{1}$}
\affiliation{$^{1}$Department of Physics and Astronomy, California State University, Northridge, California 91330, USA\\
$^{2}$Kavli Institute for Theoretical Physics, University of California, Santa Barbara, California 93106-4030, USA}

\begin{abstract}
  We study the quantum phase diagram of the spin-$1/2$ Heisenberg
  model on the kagom\'e lattice with first-, second-, and third-neighbor
  interactions $J_1$, $J_2$, and $J_3$ by means of density matrix
  renormalization group.  For small $J_2$ and $J_3$, this model
  sustains a time-reversal invariant quantum spin liquid phase.  With
  increasing $J_2$ and $J_3$, we find in addition a $q=(0,0)$ N\'{e}el
  phase, a chiral spin liquid phase, a valence-bond crystal phase, and
  a complex non-coplanar magnetically ordered state with spins forming
  the vertices of a cuboctahedron known as a \textit{cuboc1} phase.
  Both the chiral spin liquid and \textit{cuboc1} phase break time
  reversal symmetry in the sense of spontaneous scalar spin chirality.
  We show that the chiralities in the chiral spin liquid and
  \textit{cuboc1} are distinct, and that these two states are
  separated by a strong first order phase transition.  The transitions
  from the chiral spin liquid to both the $q=(0,0)$ phase and to
  time-reversal symmetric spin liquid, however, are consistent with
  continuous quantum phase transitions.
\end{abstract}

\pacs{73.43.Nq, 75.10.Jm, 75.10.Kt}
\maketitle

\section{Introduction}

Quantum spin liquids (QSLs) are highly entangled states of
matter\cite{Nature_464_199}\ with remarkable properties of great
intrinsic interest.  The simplest and perhaps most striking subclass
of QSLs comprises topologically ordered states\cite{PRB_40_7387,
  PRB_41_9377, IJMPB_4_239}, which have a non-vanishing excitation
gap, and support emergent quasiparticles with anyonic statistics and
fractional quantum numbers\cite{Science_235_1196, RMP_78_17,
  Science_321_1306, PRB_44_2664, PRB_62_7850}. Although QSLs have been
demonstrated  in many contrived models \cite{PRL_61_2376, PRL_86_1881,
  PRB_65_224412, PRB_66_205104, PRL_94_146805, AP_321_2, NP_7_772,
  PRL_99_247203}, their relation to magnetically
disordered phases in different frustrated systems remains poorly
understood.  This problem has been intensively studied in the past by
many theoretical approaches.

Apart from their intrinsic interest, motivation to understand QSL
phases comes from recent experimental discoveries.  The kagom\'e
antiferromagnets herbertsmithite and kapellasite have recently emerged
as prominent examples\cite{PRL_98_077204, PRL_98_107204,
  PRL_103_237201, PRB_82_144412, PRL_109_037208, Nature_492_7429,
  PRL_110_207208}.  The absence of magnetic order is evidenced by many
different techniques including muon spin rotation and susceptibility
measurements \cite{PRL_98_077204, PRL_98_107204, PRL_103_237201,
  PRB_82_144412, PRL_109_037208, Nature_492_7429, PRL_110_207208}, and
neutron scattering measurements of herbertsmithite show a purely
continuum spectrum, interpreted as a signature of fractional
``spinon'' excitations\cite{Nature_492_7429}.  Further instances of
QSLs have been found in organic Mott insulators with a triangular
lattice structure\cite{PRL_91_107001, PRL_95_177001, PRB_77_104413}.

Theoretically, QSLs have been sought in spin-$1/2$ antiferromagnets
with frustrated and/or competing interactions on triangular
\cite{PRL_69_2590, PRB_72_045105, PRB_74_012407, PRL_98_077205},
honeycomb \cite{PRB_82_024419, PRB_84_024420, PRL_110_127203,
  PRL_110_127205, PRB_88_165138}, square \cite{PRL_66_1773,
  PRB_86_024424, PRL_113_027201}, and kagom\'e \cite{PRL_101_117203,
  Science_332_1173, PRL_109_067201, NP_8_902, NatCom_4_2287} lattices.
Amongst all these, the kagom\'e Heisenberg model (KHM) appears to
possess the most robust QSL phase, and the only one consistently found
in unbiased density matrix renormalization group (DMRG) calculations.
The nature of the QSL is less clear.  The DMRG studies suggest a
gapped QSL \cite{PRL_101_117203, Science_332_1173, PRL_109_067201,
  NP_8_902}, seemingly consistent with $Z_2$ topological
order \cite{PRL_109_067201,NP_8_902}.  Variational studies using
projected fermionic parton wavefunctions favor a different, gapless
Dirac state \cite{PRL_98_117205, PRB_87_060405, PRB_89_020407}.  A bosonic parton
wavefunction does, however, give a competitive energy for a $Z_2$ QSL
state in the extended $J_1$-$J_2$ KHM with second-neighbor
antiferromagnetic exchange $J_2$ \cite{PRB_84_020404}, and other
studies reinforce an enhanced QSL phase in this model \cite{NP_8_902,
  EPL_88_27009, APS_White, PRB_89_020408}. Direct evidence for $Z_2$
topological order is mixed: in support, a nearly quantized
topological entanglement entropy was found in the $J_1$-$J_2$
model \cite{NP_8_902}, but the expected four topological ground state
sectors have not been seen in DMRG \cite{PRB_89_075110}.

Interestingly, by introducing {\em both} second and third neighbor
couplings, DMRG studies \cite{1312.4519, PRL_112_137202, 1407.2740,
  XY_model} recently discovered another topological QSL on the kagom\'e
lattice.  This state spontaneously breaks time reversal symmetry (TRS)
in the sense of having a complex wavefunction and non-zero scalar spin
chirality $\chi_{ijk} = S_i \cdot (S_j \times S_k)$ for some triplets of
nearby spins $i,j,k$.  Such a state, proposed more than 20 years ago
by Kalmeyer and Laughlin \cite{PRL_59_2095, PRB_39_11413},
is known as a Chiral Spin Liquid (CSL).  It can be regarded as a spontaneous
fractional quantum Hall effect.  The CSL occurs in several different
kagom\'e spin models with comparable $J_2$ and $J_3$\cite{
1312.4519,1407.0869, 1411.1327}, and indeed is {\em
  more} robust than the putative $Z_2$ QSL state discussed earlier:
all the expected universal topological properties of the CSL state
have been verified numerically.

\begin{figure}
\includegraphics[width=1.0\linewidth]{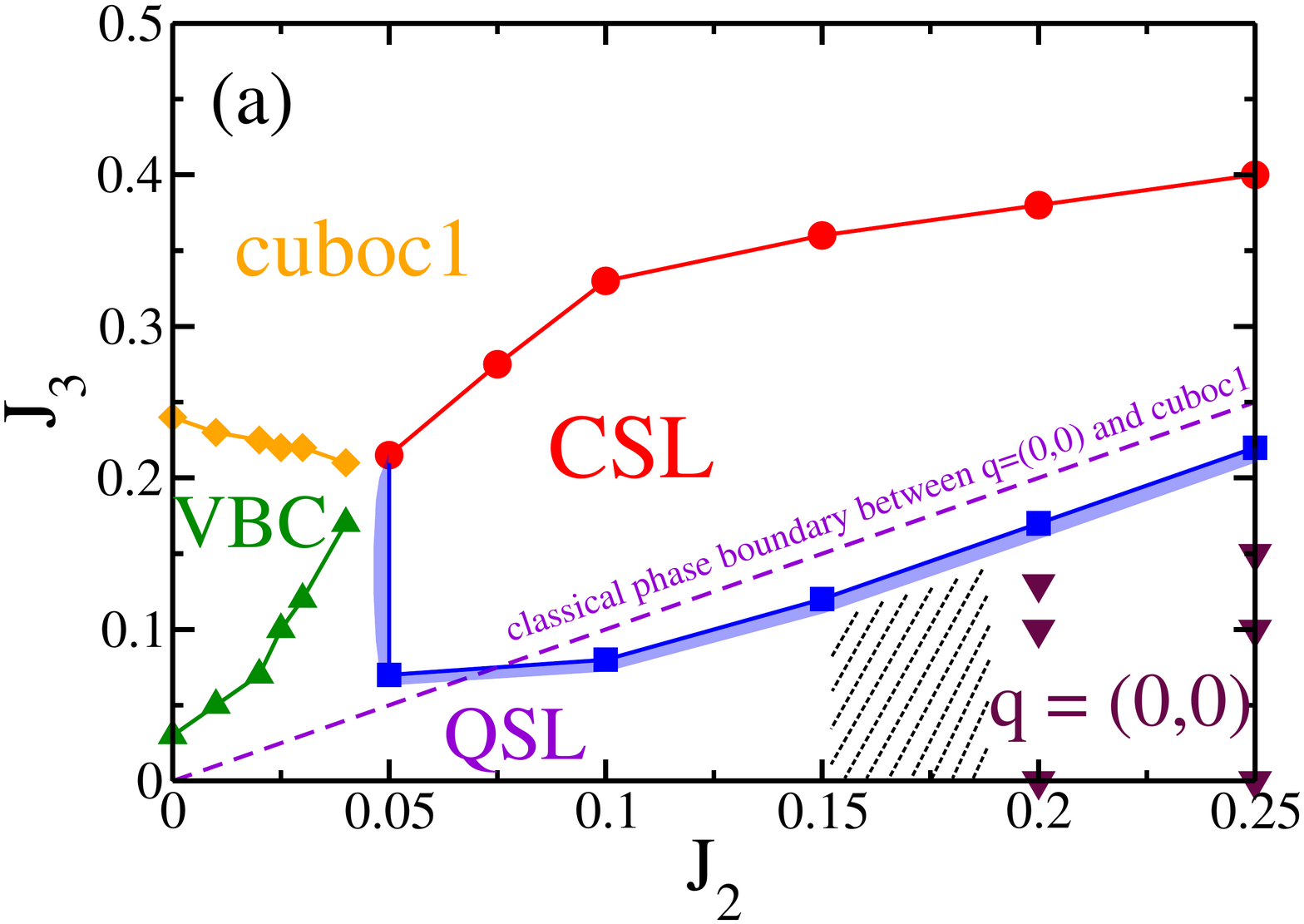}
\includegraphics[width=0.8\linewidth]{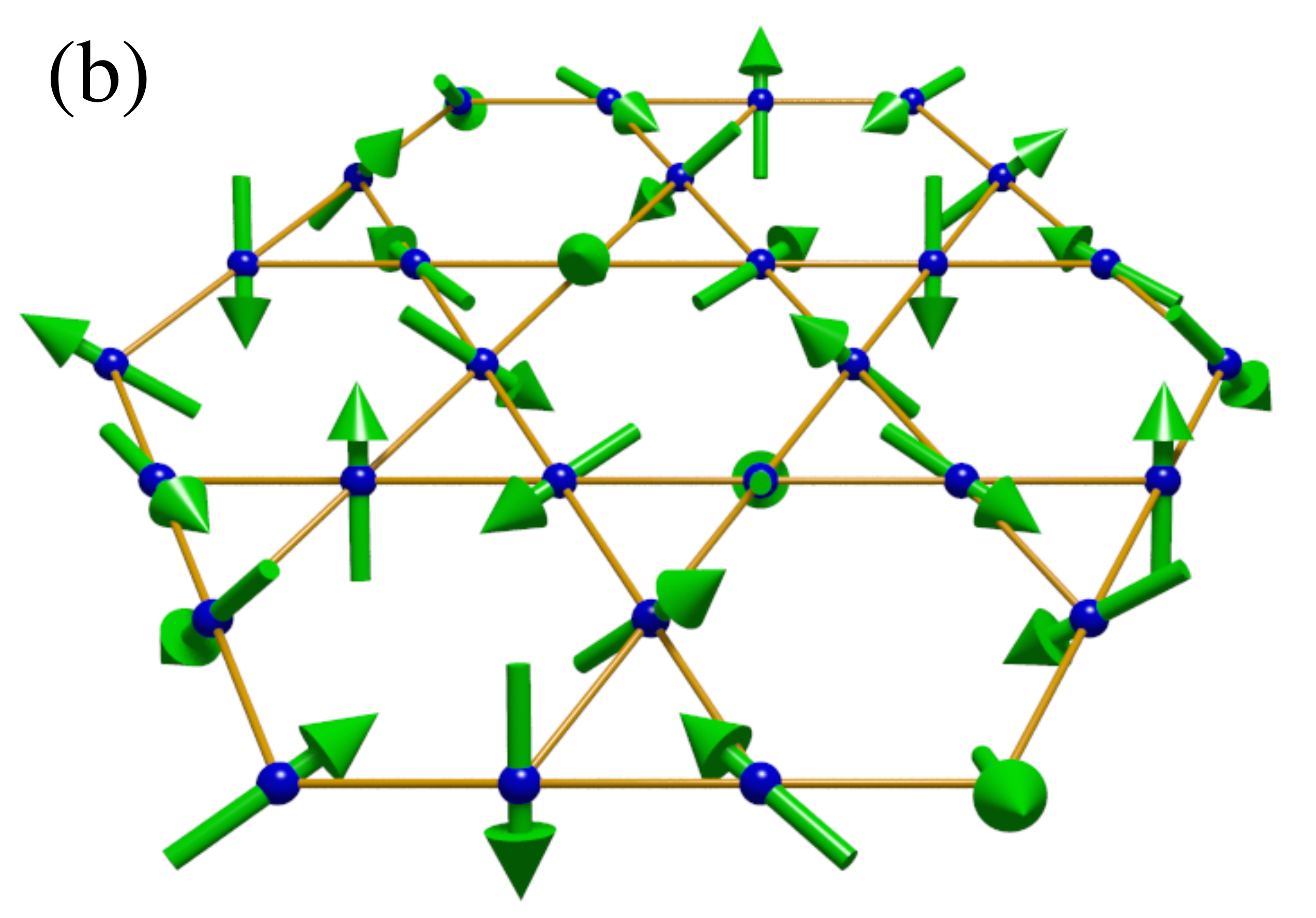}
\caption{(a) Quantum phase diagram of the spin-$1/2$ $J_1$-$J_2$-$J_3$
  kagom\'e Heisenberg model for $0.0 \leq J_2 \leq 0.25$ and
  $0.0 \leq J_3 \leq 0.5$.  The phases shown are: a time-reversal
  invariant quantum spin liquid (QSL) phase, a coplanar magnetically
  ordered $q = (0,0)$ N\'{e}el phase, a time-reversal broken chiral
  spin liquid (CSL) phase, a non-coplanar magnetically and chiral
  ordered \textit{cuboc1} phase, and a valence bond crystal (VBC)
  phase.  The \textit{cuboc1} phase remains stable for larger $J_3$
  beyond the range shown here (we have checked up to $J_3 \leq 1.0$).
  The dashed region indicates the uncertainty in locating the phase
  boundary between the QSL and $q=(0,0)$ N\'eel phases.  The purple
  dashed line shows the line of classical degeneracy between the
  $q=(0,0)$ N\'eel and \textit{cuboc1} phases\cite{PRL_108_207204}.
  (b) The configurations of spins (arrows indicate the direction of
  static moments) of the \textit{cuboc1} state on the kagom\'e
  lattice.  On each small triangle, the spins are coplanar and sum to
  zero. In each hexagon, sets of three consecutive spins are
  non-coplanar, as are the sets obtained by taking every second spin
  around the hexagon.  This breaks time-reversal symmetry in the sense
  that the scalar spin chirality is non-zero and the wavefunction is
  intrinsically complex.}\label{phase}
\end{figure}

In this paper, we expose the relations between the two QSL states and
nearby ordered phases through a global DMRG\cite{PRL_69_2863} study of the full phase diagram
of the $J_1$-$J_2$-$J_3$ model (with all exchanges antiferromagnetic):
\begin{equation}
H = J_1 \sum_{\langle i,j\rangle}S_i\cdot S_j + J_2 \sum_{\langle\langle i,j\rangle\rangle}S_i \cdot S_j + J_3 \sum_{\langle\langle\langle i,j\rangle\rangle\rangle}S_i \cdot S_j. \label{Hamiltonian}
\end{equation}
Good points of comparison are the classical and
Schwinger boson mean field phase diagrams, found in
Ref. \onlinecite{PRL_108_207204}.  These studies found two magnetically
ordered phases breaking TRS, known as \textit{cuboc1} and
\textit{cuboc2}, as well as a simpler $q=(0,0)$ coplanar ordered state
which is time-reversal symmetric.  They also conjectured that a $Z_2$
TRS breaking QSL ``descended'' from the \textit{cuboc1} state by
quantum disordering of the spins might apply to the the pure
nearest-neighbor KHM, and also extend to the region with small $J_2$
and $J_3$ perturbations \cite{PRL_108_207204}.  The DMRG phase diagram
determined here is shown in Fig.~\ref{phase}(a), and bears out some
but not all of these features.  We indeed find the ordered \textit{cuboc1}
(see Fig.~\ref{phase}(b) of the spin configuration of \textit{cuboc1} state)
and $q=(0,0)$ states when $J_3$ or $J_2$ are large, roughly correlating
with their classical positions.  These classical states surround three
more quantum ones: the two aforementioned QSL states and a Valence
Bond Crystal (VBC) state with ordered singlets breaking translational
but not spin-rotation or TRS symmetry. The relations between these
states and the classical ones will be discussed below.

\begin{figure}
\includegraphics[width=0.495\linewidth]{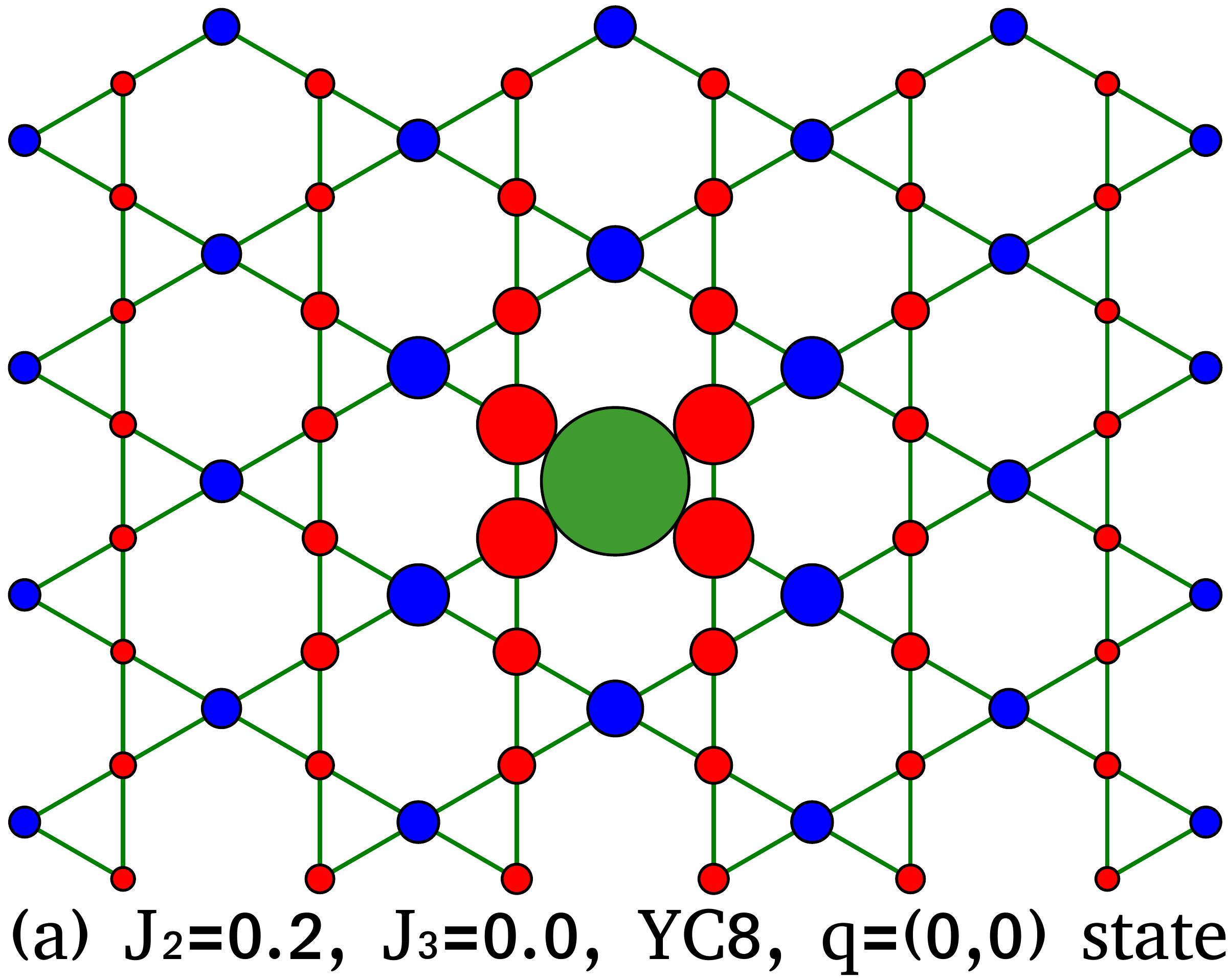}
\includegraphics[width=0.495\linewidth]{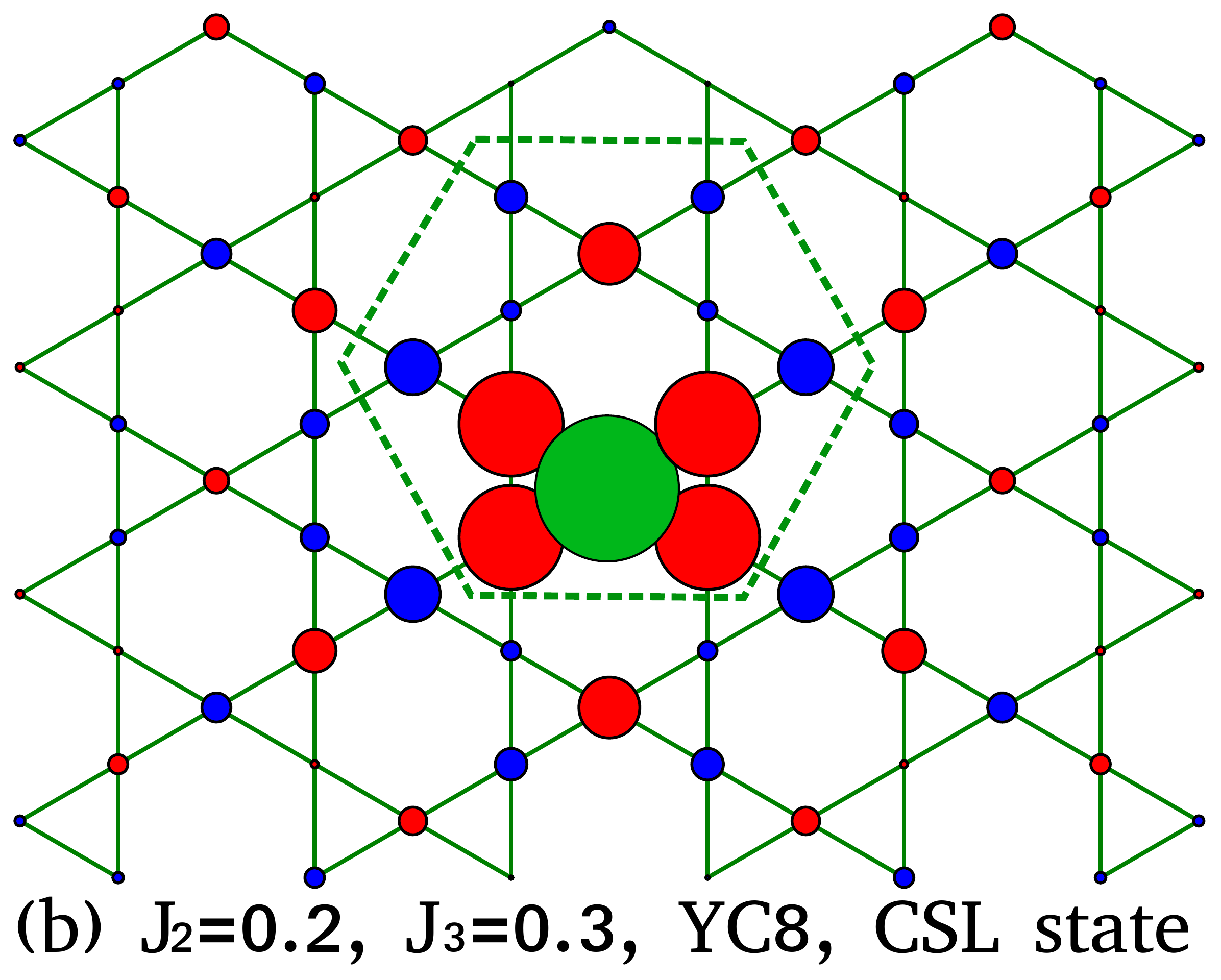}
\includegraphics[width=0.495\linewidth]{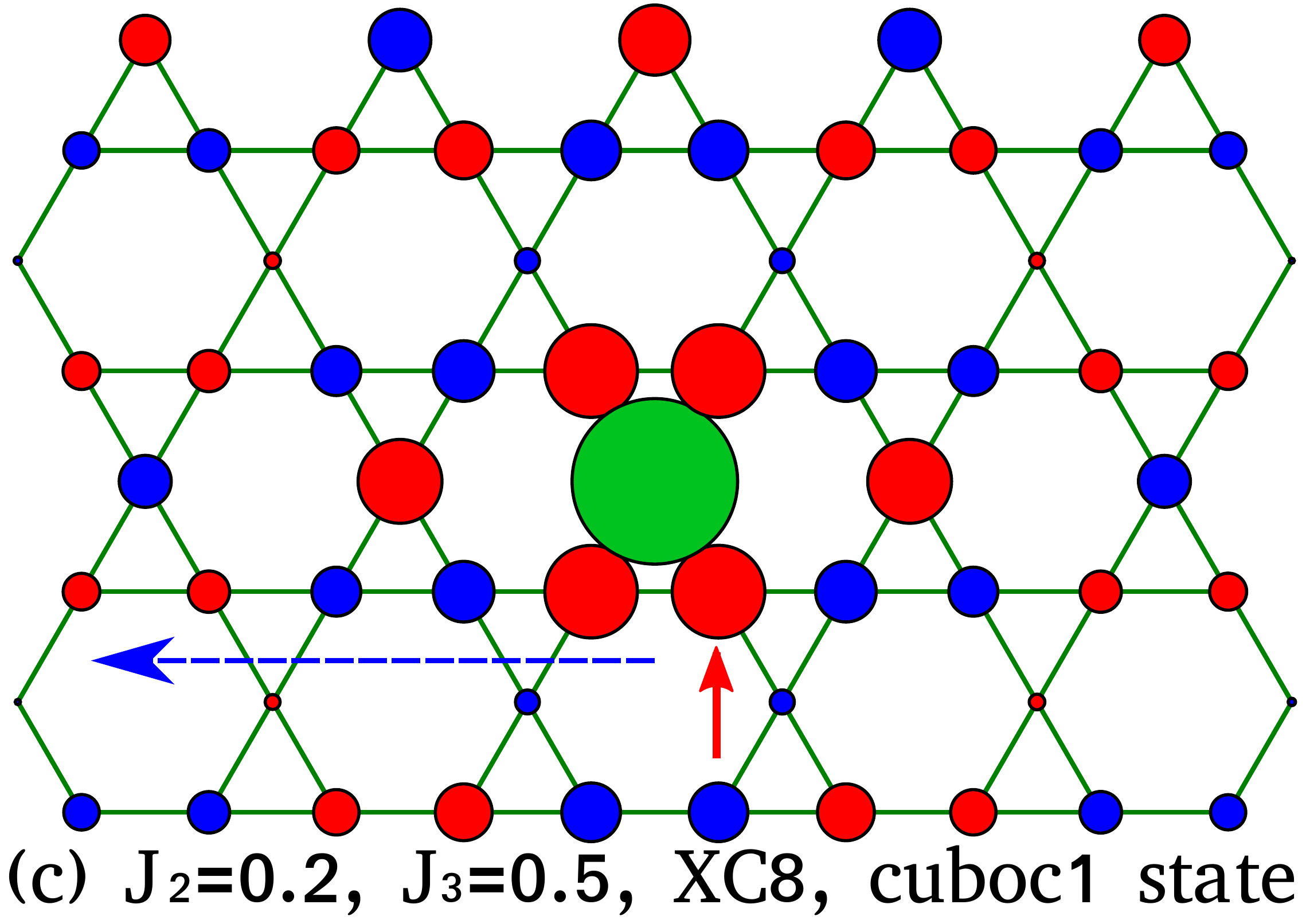}
\includegraphics[width=0.495\linewidth]{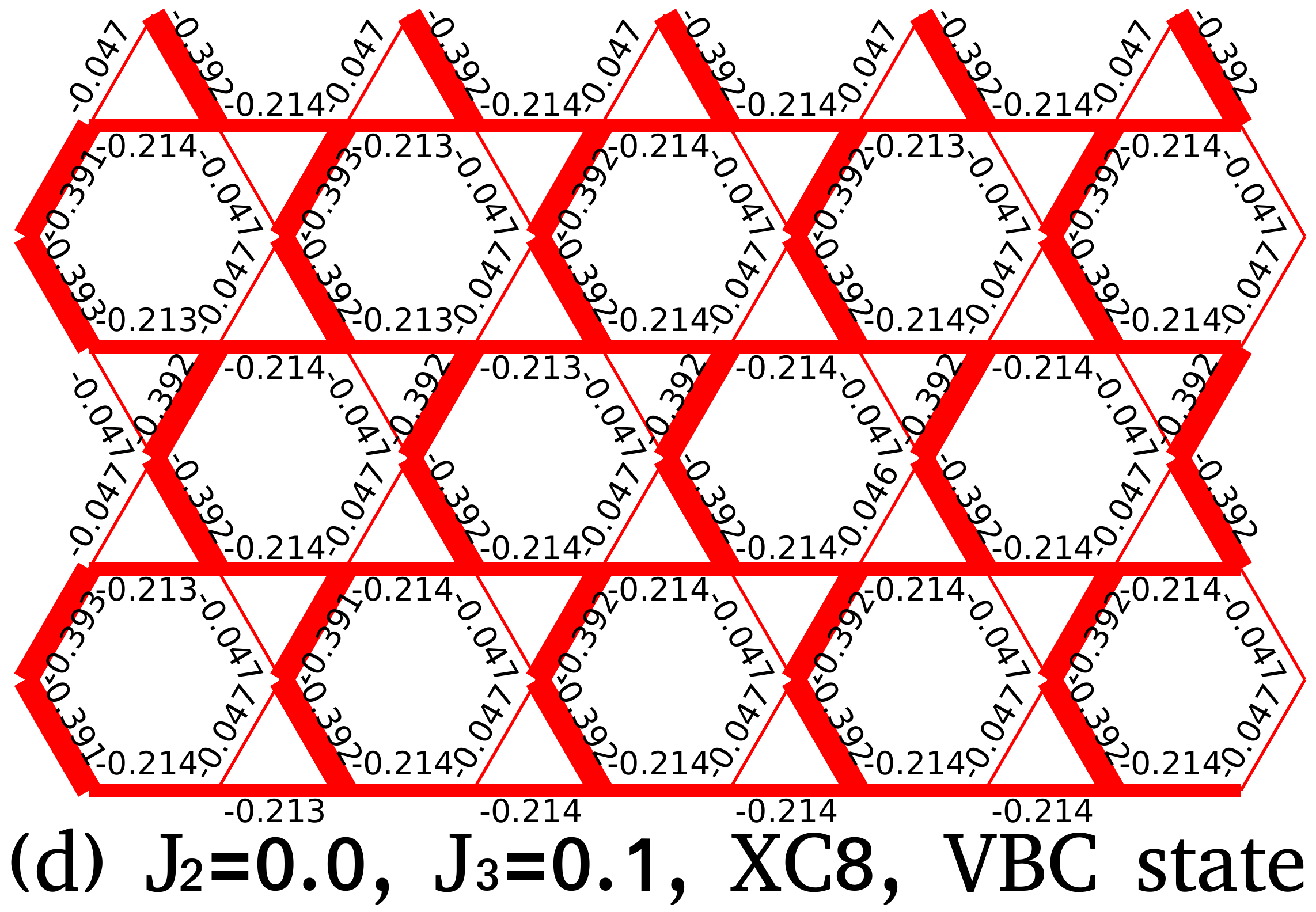}
\caption{(a)-(c) show
  the spin-spin correlations for different phases on the YC8 and XC8
  cylinders. The green site is the reference spin, the blue and red
  colors denote positive and negative correlations, respectively, of
  the site in question with the reference spin.  The area of circle is
  proportional to the magnitude of the spin correlation. The large dashed
  hexagon in (b) shows the short-range spin correlations in CSL phase. The arrows in
  (c) show the reference spin (the red solid arrow) and the direction
  of other spins (the blue dashed arrow) whose correlations are
  plotted in Fig.~\ref{j1j2}.  Panel (d) plots the nearest-neighbor
  bond energy on the XC8 cylinder in the VBC phase. The same pattern
  is observed on the XC12 cylinder in the VBS phase.}\label{pattern}
\end{figure}

For this study, we use the DMRG with $SU(2)$ spin rotational symmetry
\cite{EPL_57_852} on cylinders by keeping a number of $U(1)$-equivalent states $M$ as
large as $M_{\rm max}=26000$.  Two cylinder geometries, denoted XC and YC,
are used, such that for the XC (YC) cylinder, one of the three bond
orientations is along the $x$ ($y$) axis, as shown in
Fig.~\ref{pattern}.  We abbreviate specific cylinders
by XC$2L_y$-$L_x$ and YC$2L_y$-$L_x$, where $L_x$ ($L_y$) is the number
of unit cells in the $x$ ($y$) direction.  In general, we obtain
results with DMRG truncation error less than $1\times 10^{-6}$ and
$1\times 10^{-5}$ for the cylinders with $L_y = 4$ and $6$, respectively.

\section{$q=(0,0)$ N\'{e}el phase in the $J_1$-$J_2$ KHM}

We begin by studying the $q=(0,0)$ N\'{e}el order in the small $J_2$
region with $J_3=0$, and first investigate the spin correlations on
cylinders of varying widths.  A gapped magnetically disordered phase
would be expected to show exponentially decaying correlations.  In a
long-range magnetically ordered phase, the correlations should remain
non-zero in magnitude at long distances {\em in two dimensions}.  On a
long cylinder of even width, exponential decay is still expected even
when the two dimensional limit is ordered, but in that case the decay
is characterized by a correlation length $\xi$ which grows linearly with system
width. Thus it is crucial to investigate the scaling of the correlation length.

\begin{figure}
\includegraphics[width=0.51\linewidth]{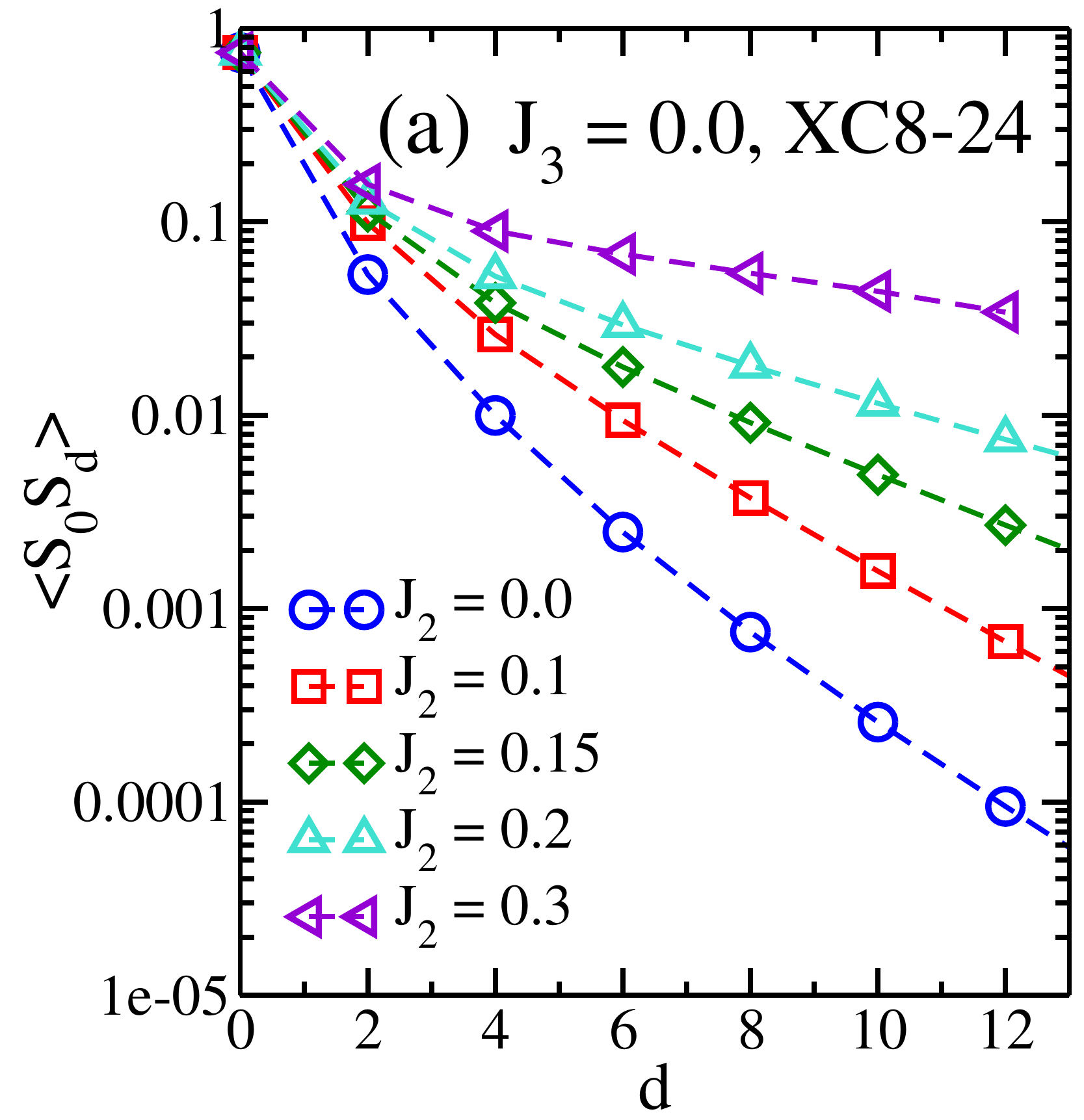}
\includegraphics[width=0.475\linewidth]{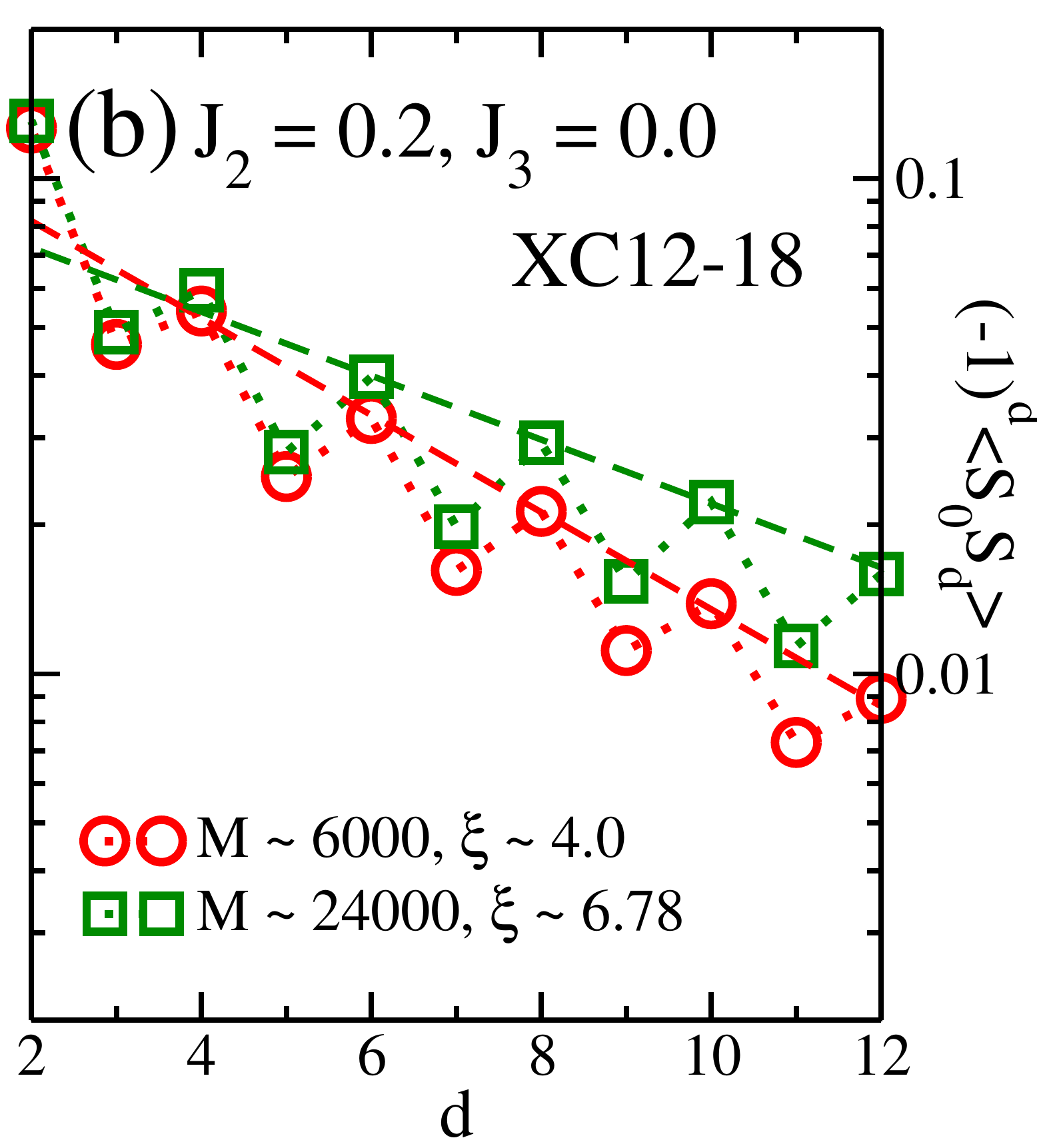}
\includegraphics[width=0.51\linewidth]{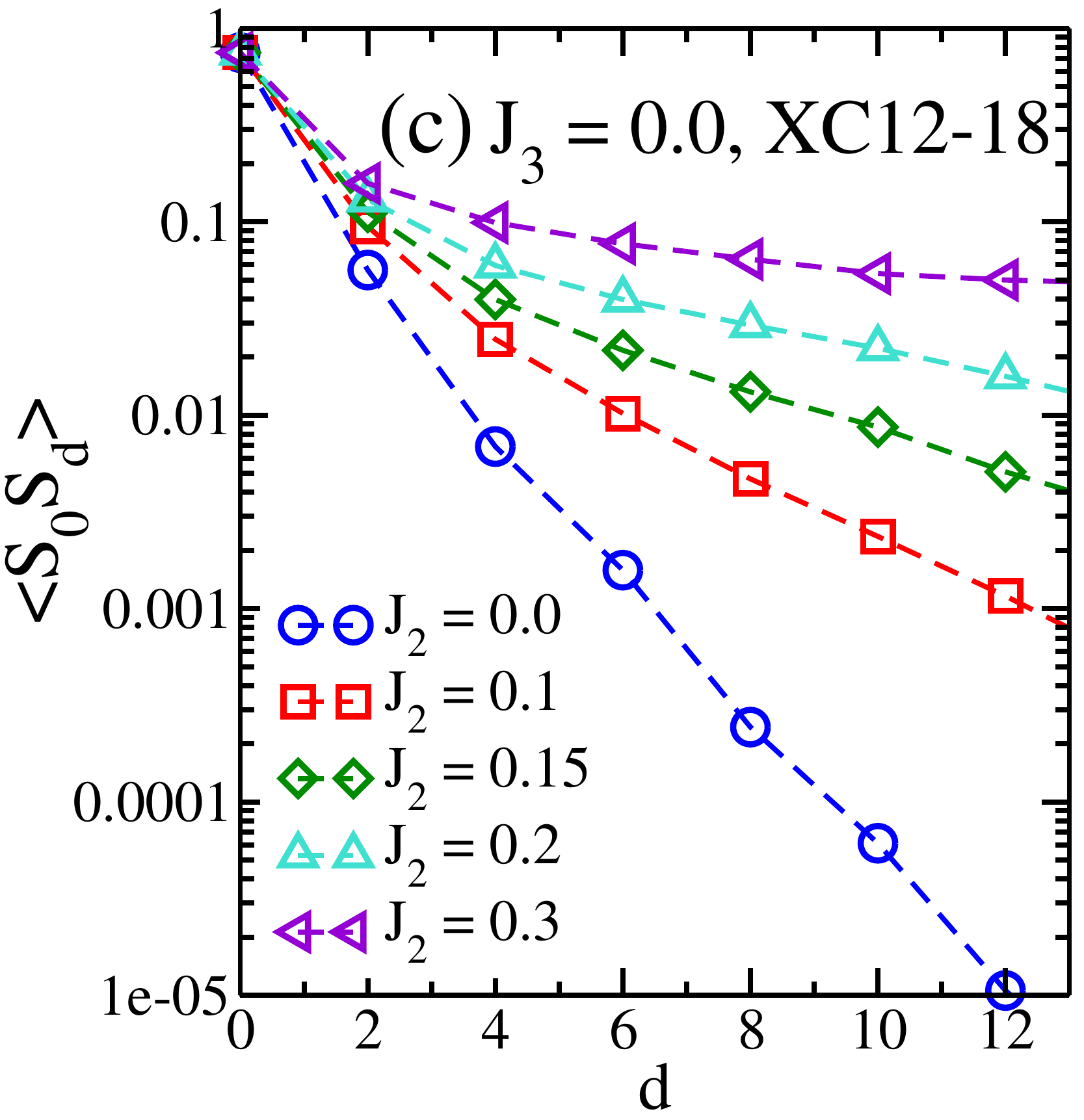}
\includegraphics[width=0.48\linewidth]{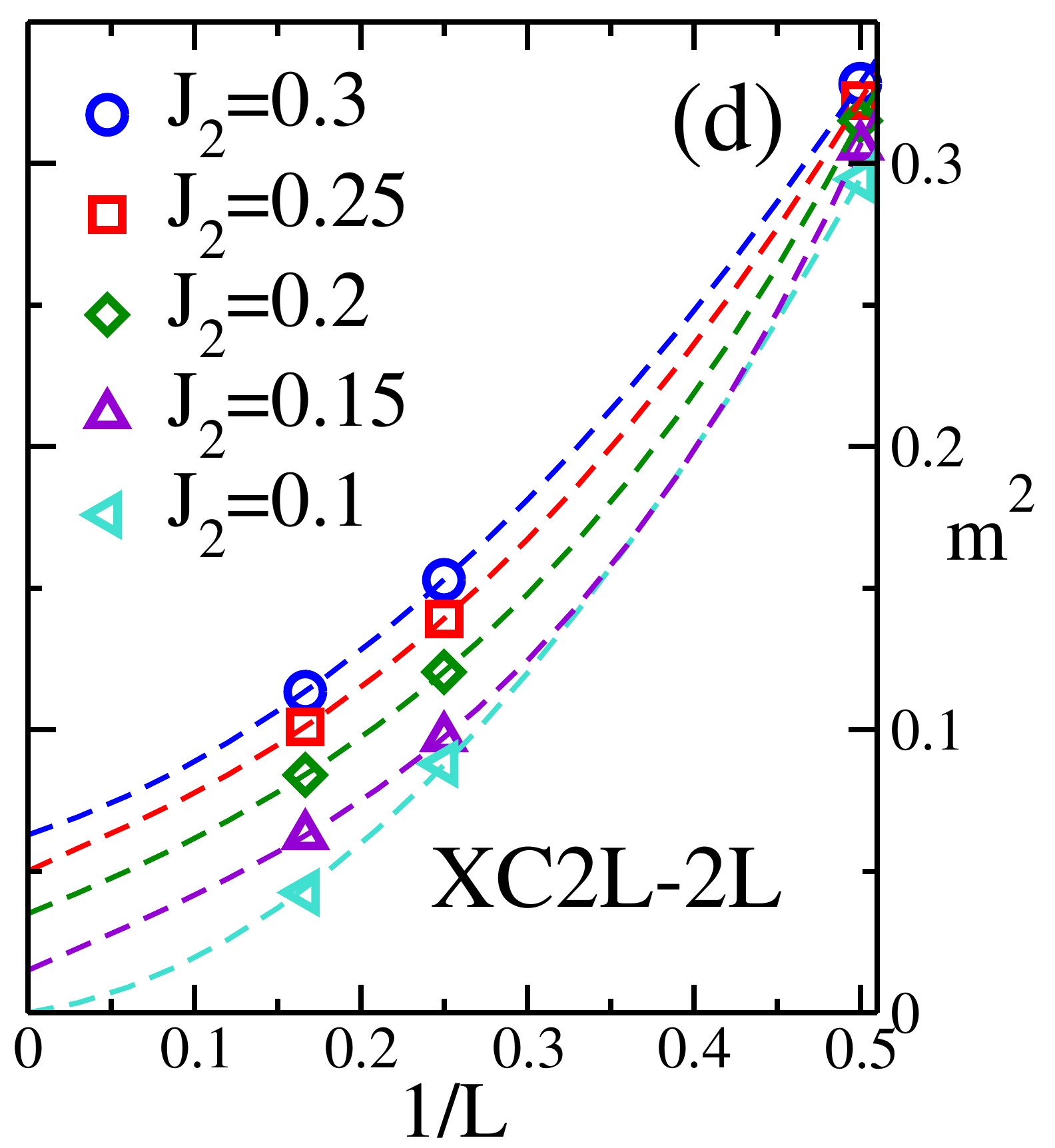}
\caption{Panels (a) and (c) are log-linear plots of the spin
  correlations between sites of one sublattice versus site distance
  $d$ on the XC8 and XC12 cylinders. Panel (b) illustrates the
  dependence upon the number of states kept, $M$, in the DMRG, for the
  case $J_2 = 0.2$ on the XC12 cylinder.  The geometry of the sites in
  the first three panels is shown in Fig. \ref{pattern}(c). Panel (d)
  shows the finite size scaling of the (squared) magnetic order
  parameter $m^2$ versus $1/L$ on the XC4, XC8, and XC12
  cylinders. All plots in this figure are for $J_3=0$. }\label{j1j2}
\end{figure}

Fig.~\ref{j1j2}(a) shows the correlations between spins on the same
sublattice, $\langle S_0 \cdot S_d \rangle$, on the XC8-24 cylinder.
One sees that the spin correlation length continues to grow with
increasing $J_2$.  At $J_2 =0.2$, the system appears to develop
longer-range correlation.  While the results on the XC8 cylinder are
fully converged, those on the wider XC12 cylinder are not, and display
dependence on the number of states kept in DMRG.  In particular,
$\xi$ grows as more states are included.  Therefore we measure the
spin correlations for several different numbers, $M= 4000 - 24000$, of
$U(1)$-equivalent states and extrapolate the result. For example at
$J_2 = 0.2$ (shown in Fig.~\ref{j1j2}(b)), $\xi$ is $4.0$ lattice
spacings for $M\approx 6000$ states while it grows to $6.78$
lattice spacings when $M\approx 24000$ states.  Thus, the less
converged results may underestimate N\'{e}el order.  In
Fig.~\ref{j1j2}(c), we show the spin correlations on the XC12
cylinder obtained with $M \approx 24000$ states.  Comparing to
Fig.~\ref{j1j2}(a), we observe that for $J_2=0$, the correlation
length {\em decreases} with cylinder width, while it {\em increases}
with width for $J_2 = 0.1 - 0.2$.  The trend is at least indicative of
growing order.

Next, we study the $J_2$ dependence of the magnetic order parameter,
extrapolating to the thermodynamic limit from finite-size systems.  We
calculate the finite-size order parameter from the middle half of each
system to minimize boundary effects, i.e. the $3 \times L \times L$
sites out of a total of $3 \times L \times 2L$ sites for the
XC$2L$-$2L$ ($L = 2,4,6$) cylinders.
The order parameter of the $q=(0,0)$ N\'{e}el state is defined as
$m^2 = \frac{1}{N^2} \sum_{i,j} \langle S_i \cdot S_j \rangle$ ($N$ is
the number of unit cells, and $i,j$ are sites in the same sublattice).
In Fig.~\ref{j1j2}(d), we show $m^2$ versus $1/L$ for various $J_2$ \cite{convergence}.
For $J_2 \geq 0.15$, we find that $m^2$ extrapolates to finite values in
the thermodynamic limit, indicative of $q=(0,0)$ N\'{e}el long range
order.  Because of the limited range of system width, we have
considerable uncertainty in the location of the phase boundary and
cannot reliably estimate an error bar. However, we feel confident that the magnetic order is
robust for $J_2=0.2$, as shown in Fig.~\ref{j1j2}(b-c), which sets a
lower bound on the transition point.


\section{Chiral spin liquid phase}

\begin{figure}
\includegraphics[width=0.85\linewidth]{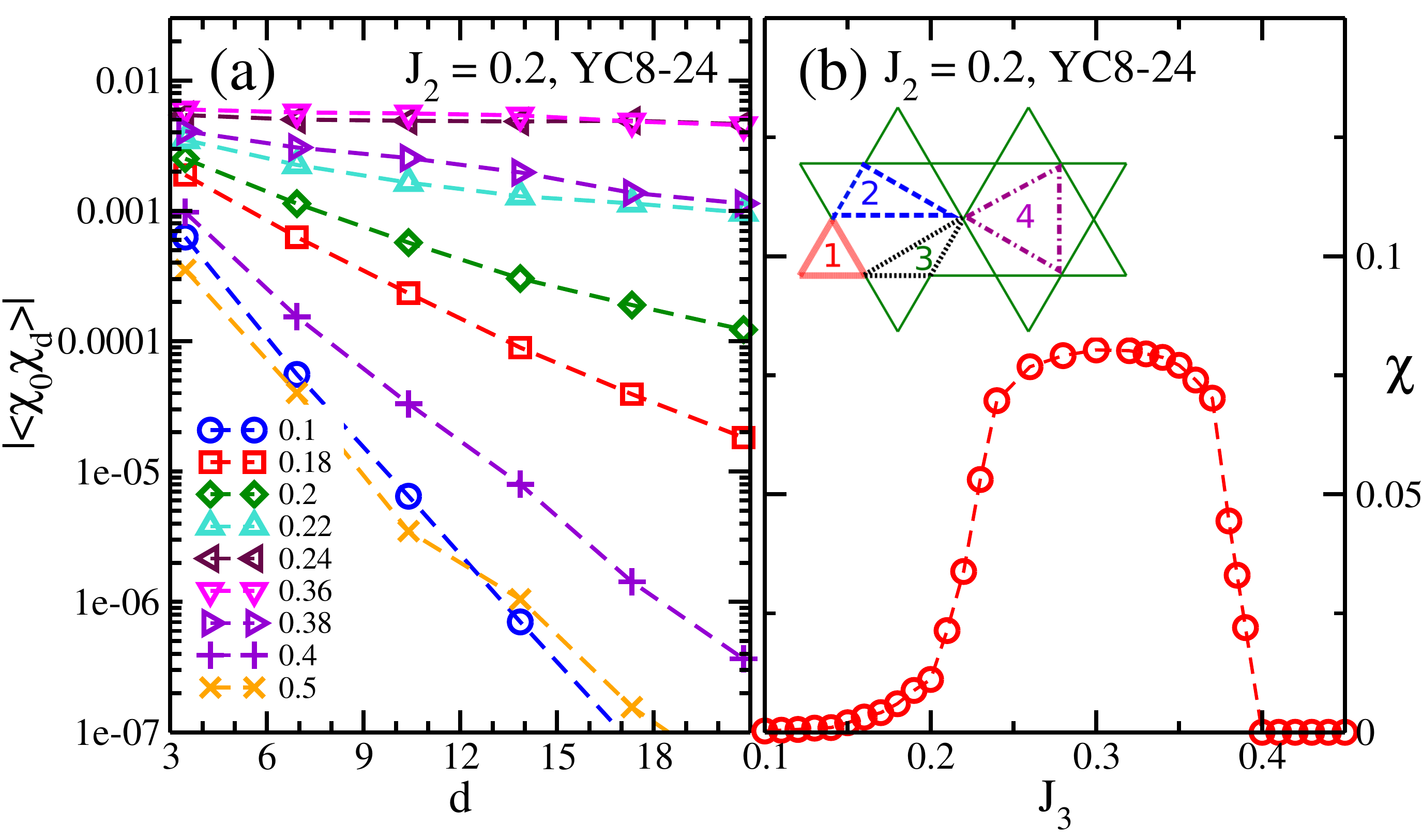}
\includegraphics[width=0.8\linewidth]{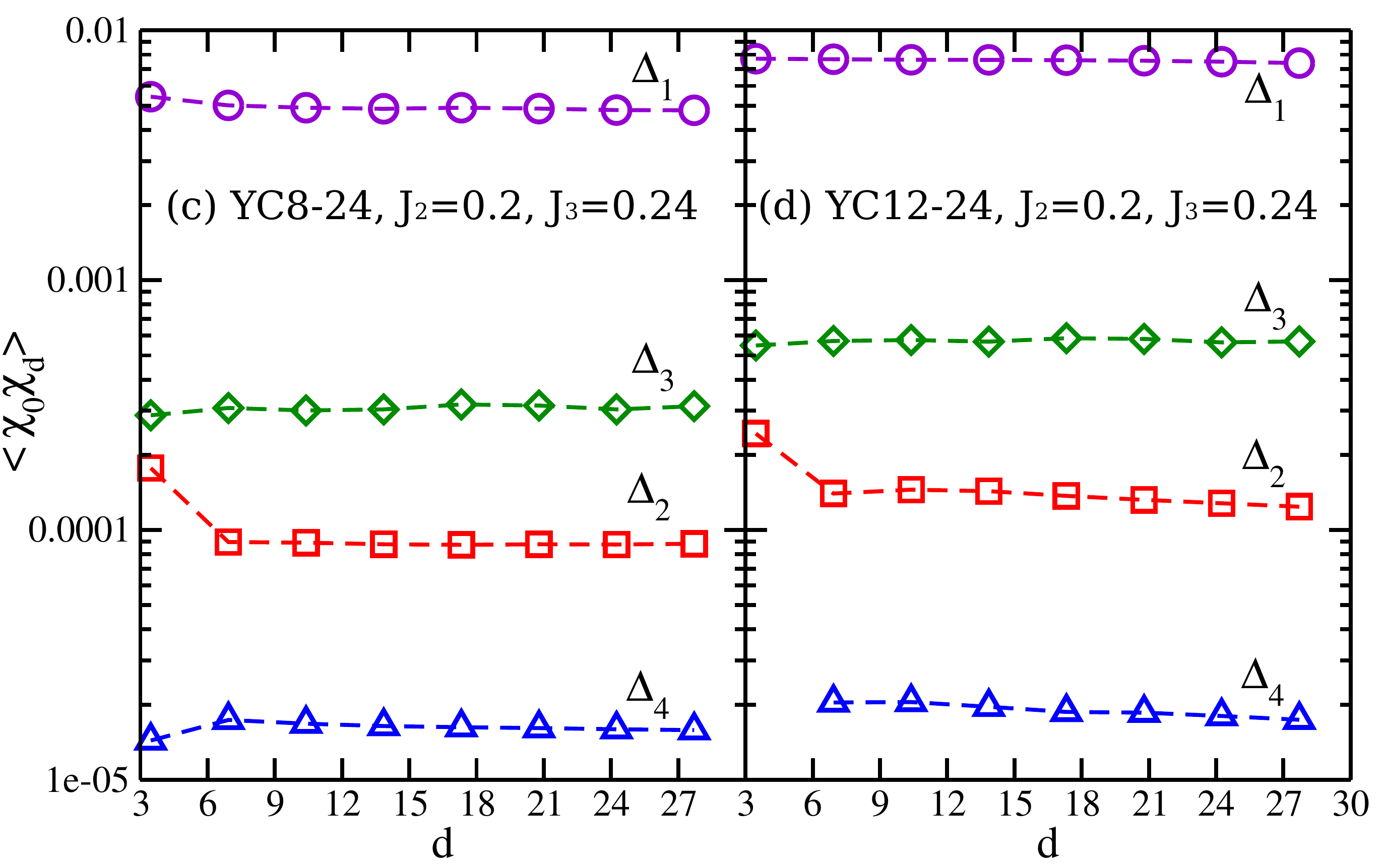}
\includegraphics[width=0.6\linewidth]{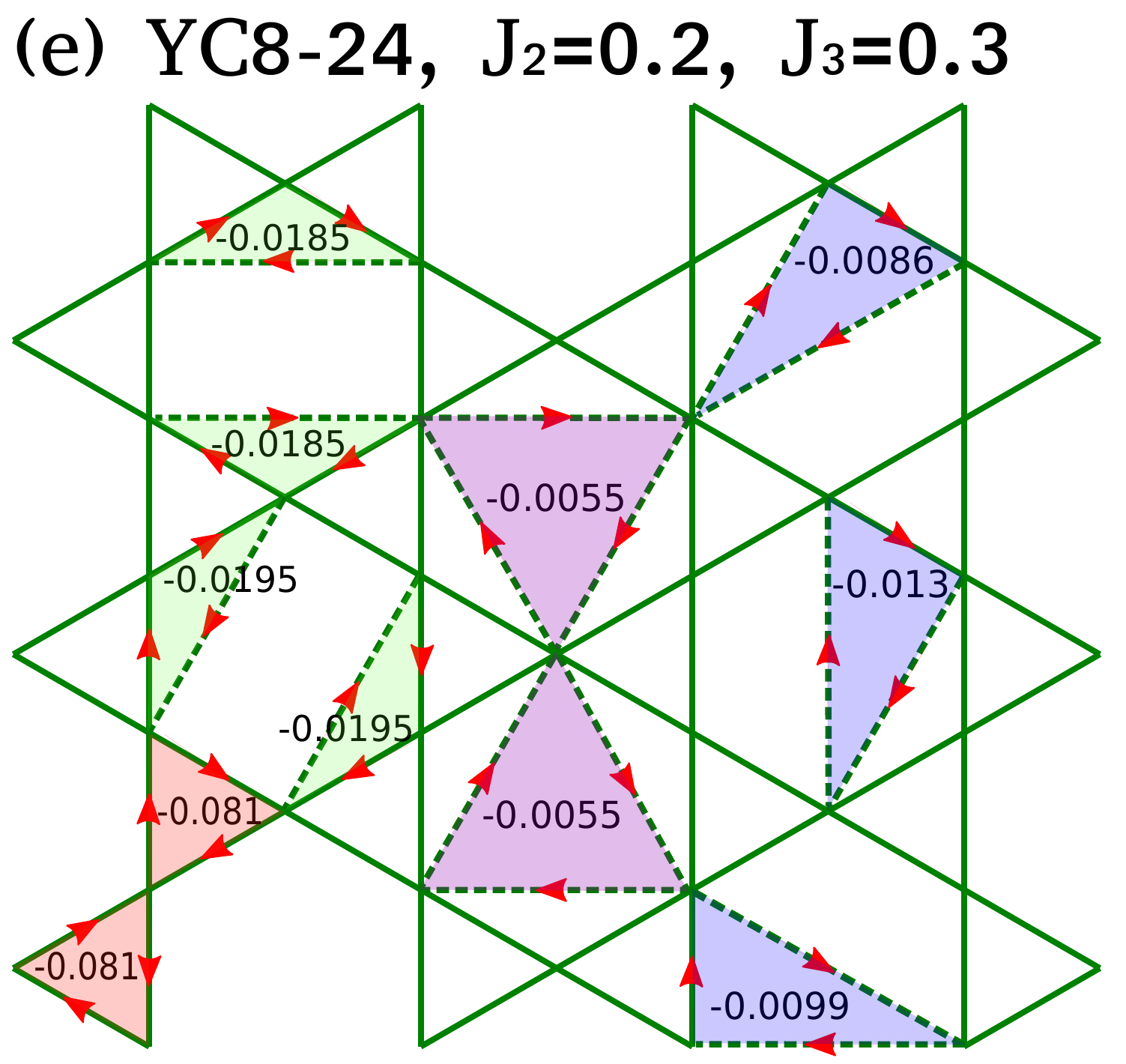}
\caption{Numerical results for scalar spin chirality.  Panel (a) shows
  correlations between pairs of the smallest triangles $\Delta_1$ on
  the kagom\'e lattice for $J_2 = 0.2$ and different values of $J_3$
  on the YC8-24 cylinder. In (b), the long-distance value of these
  correlations is extracted and plotted versus $J_3$, which clearly
  shows the CSL region.  Plots (c) and (d) compares the correlations
  of different types of triangles (defined in the inset of (b)) for
  $J_2 = 0.2, J_3 = 0.24$ on the YC8-24 and YC12-24 cylinders.  Panel
  (e) shows the chiral order parameter
  $\langle \chi_{\Delta_i} \rangle$ calculated directly from the
  complex code for the indicated triangles on the YC8-24 cylinder at
  $J_2 = 0.2, J_3 = 0.3$. The red arrows indicate the order of the
  three spins in the triple product defining the scalar spin
  chirality.  }\label{chiral}
\end{figure}

Prior work has fully established the CSL state in the
$J_1$-$J_2$-$J_3$ KHM along the parameter line
$J_2 = J_3 = J^{\prime}$ with
$0.1 \lesssim J^{\prime} \lesssim 0.7$\cite{1312.4519}. Moreover, the
topological order of the CSL was found to be that of the $\nu = 1/2$
Laughlin state\cite{1312.4519}.  This completely fixes the universal
topological aspects of the CSL.  Here we study some non-universal
aspects of the CSL which help to show its relation to the surrounding
phases.  First, we determine the complete domain of the CSL phase
through a study of the scalar spin chirality.  In
Fig. \ref{chiral}(a), we show the correlation function between
chirality on pairs of the smallest triangles of the kagom\'e lattice
(indicated with the number ``1'' in the inset of
Fig. \ref{chiral}(b)), for the YC8-24 cylinder with $J_2 = 0.2$ and
various $J_3$. In the $q=(0,0)$ N\'{e}el phase, for example for
$J_3 = 0.1$, the chiral correlations decay rapidly and exponentially
to zero.  With increasing $J_3$, the chiral correlations gradually
grow, apparently establishing long-range order (i.e. saturating to a
finite value at large distance) at $J_3 \simeq 0.22$.  This behavior
persists to $J_3 \simeq 0.4$, beyond which the chiral correlations
exhibit a sharp decrease. We define $\chi$ as the square root of the
long-distance chiral correlations in Fig. \ref{chiral}(a)
$\chi \equiv \sqrt{|\langle \chi_0 \chi_d \rangle |}$ ($d$ is the
longest available distance) to describe the variation of chiral
correlation. Fig.~\ref{chiral}(b) shows the $J_3$ dependence of
$\chi$, which clearly indicates the CSL phase exists over a
well-defined but limited range of $J_3$.  From this figure, we
conservatively estimate $\chi=0$ for $J_3 \lesssim 0.2$ and
$J_3 \gtrsim 0.4$. We note that chiral order, which breaks the
discrete $Z_2$ time reverse symmetry, can exist even in a
one-dimensional system.  So we must carefully consider the behavior on
wider cylinders to firmly establish the presence of chiral order in
{\em two} dimensions.  To do so, we compare the behavior on the YC8
cylinder to that on XC12 and YC12 cylinders, as in
Figs.~\ref{chiral}(c) and \ref{chiral}(d). We see that for some exchange parameters, the
chiral order grows {\em stronger} with increasing system width, which
we take as evidence for time-reversal symmetry breaking in two dimensions.  Using this behavior as
a first criterion for the CSL, and the second that magnetic correlations are
short-ranged, we arrive at the shaded boundary between the $q=(0,0)$
N\'{e}el phase and the CSL shown in Fig.~\ref{phase}(a).

To further reveal the structure of the chirality in the CSL, we study
the chiral correlations between pairs of triangles of each of the four
types shown in the inset of Fig.~\ref{chiral}(b).  Results for
$J_2 = 0.2, J_3 = 0.24$ on YC8-24 and YC12-24 cylinders are shown in
Figs.~\ref{chiral}(c) and \ref{chiral}(d).  Clearly correlations of all four types of
triangles have long-range order, which demonstrates spontaneous scalar
spin chirality on all triangles. The largest chirality occurs on the
smallest triangles (labeled $\Delta_1$).
As a check, we also calculate the expectation value of the local scalar spin chirality
directly using a complex code, which allows broken time-reversal
symmetry.  The results for the YC8-24 cylinder at
$J_2 = 0.2, J_3 = 0.3$ are shown in Fig.~\ref{chiral}(e).  We see that
the expectation values are indeed all non-zero and of a uniform sign.  Moreover the magnitudes of the spontaneous
spin chirality obtained in this way obey
$|\langle \chi_{\Delta_1} \rangle| > |\langle \chi_{\Delta_3} \rangle|
> |\langle \chi_{\Delta_2} \rangle| > |\langle \chi_{\Delta_4}
\rangle|$,
consistent with the results of the correlation function analysis.  We
note that, up to very small discrepancies which we attribute to the
boundary effects due to the cylinder geometry, the spontaneous
chiralities respect the translational and rotational symmetries of the
lattice.  

\begin{figure}
\includegraphics[width=1.0\linewidth]{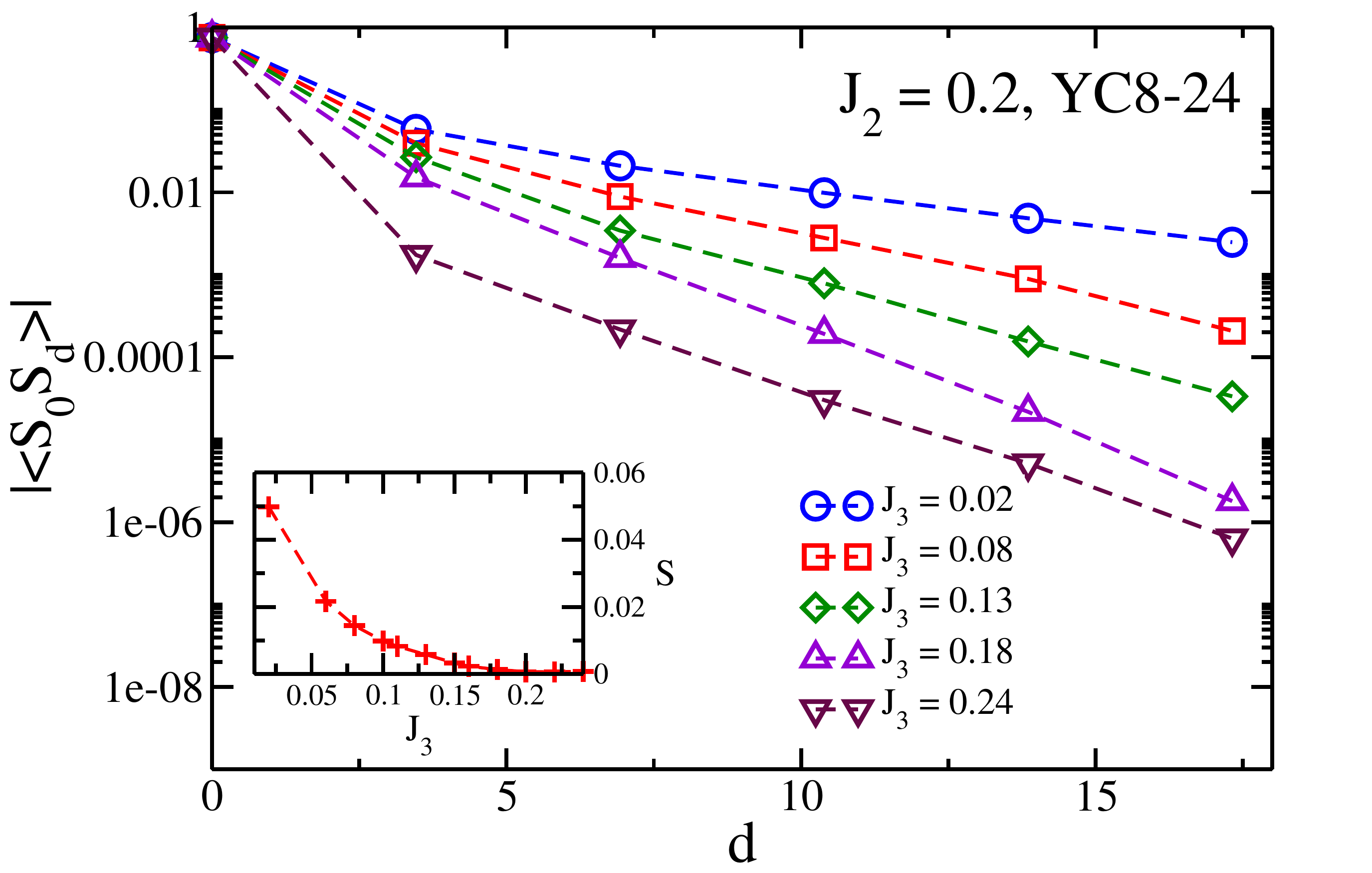}
\caption{Log-linear plot of spin correlations for $J_2 = 0.2$ on the
  YC8-24 cylinder.  These correlations decay more quickly with $J_3$
  -- a consequence of the transition from the $q=(0,0)$ N\'{e}el to
  the CSL phase. This is seen more clearly from the plot of the
  long-distance spin correlation $S$ versus $J_3$, shown in the
  inset.}\label{spin}
\end{figure}

Finally, we consider the spin correlations on passing between the
$q=(0,0)$ phase and the CSL state.  We take $J_2 = 0.2$ as an example --
see Fig.~\ref{spin}. When $J_3$ is small, the system is in the
$q=(0,0)$ phase and the spin correlations are large and slowly
decaying with a correlation length that grows with system width.  With
increasing $J_3$, the spin correlations decrease gradually.  We define
the long-distance spin correlation
$S \equiv \sqrt{|\langle S_0 \cdot S_d\rangle|}$ ($d$ is the longest
distance) as a crude estimate of magnetic order. The inset of
Fig.~\ref{spin} shows the $J_3$ dependence of $S$, which decreases
rather smoothly and for practical purposes vanishes around
$J_3 = 0.2$.  This corresponds to the onset of the CSL phase. For
larger $J_3$, the short range $q=(0,0)$ spin correlation pattern is
destroyed and the system shows instead a pattern of spin correlations
which at short distances is consistent with that of the
\textit{cuboc1} state. One such an example for $J_2 = 0.2, J_3 = 0.3$
is shown in Fig.~\ref{pattern}(b).


\section{\textit{Cuboc1} phase}
\label{sec:textitcuboc1-phase}

The \textit{cuboc1} state was first proposed for a kagom\'e
antiferromagnet in an exact diagonalization study of the $J_1$-$J_3$
KHM for $J_3 \gtrsim 0.25$\cite{JPCS_145_012008}.  It is
characterized by a $12$-sublattice non-coplanar magnetic ordering in
which the spins point towards the corners of a cuboctahedron (see Fig.~\ref{phase}(b)), 
one of the archimedean solids\cite{PRL_108_207204}.  In the classical
$J_1$-$J_2$-$J_3$ KHM, the \textit{cuboc1} phase occurs for
$J_3 > J_2$ ($J_2 < 1.0$), and shares a direct phase boundary with the
$q=(0,0)$ N\'{e}el phase as shown in Fig.~\ref{phase}(a)
\cite{PRL_108_207204}.  Owing to its non-coplanarity, the
\textit{cuboc1} state breaks time-reversal symmetry and it is natural
therefore to imagine it may be the classical ancestor of a CSL state.
Here we investigate this possibility in more detail, and argue that
the CSL in the KHM is {\em not} the descendent of the \textit{cuboc1}
state.

\begin{figure}
\includegraphics[width=0.8\linewidth]{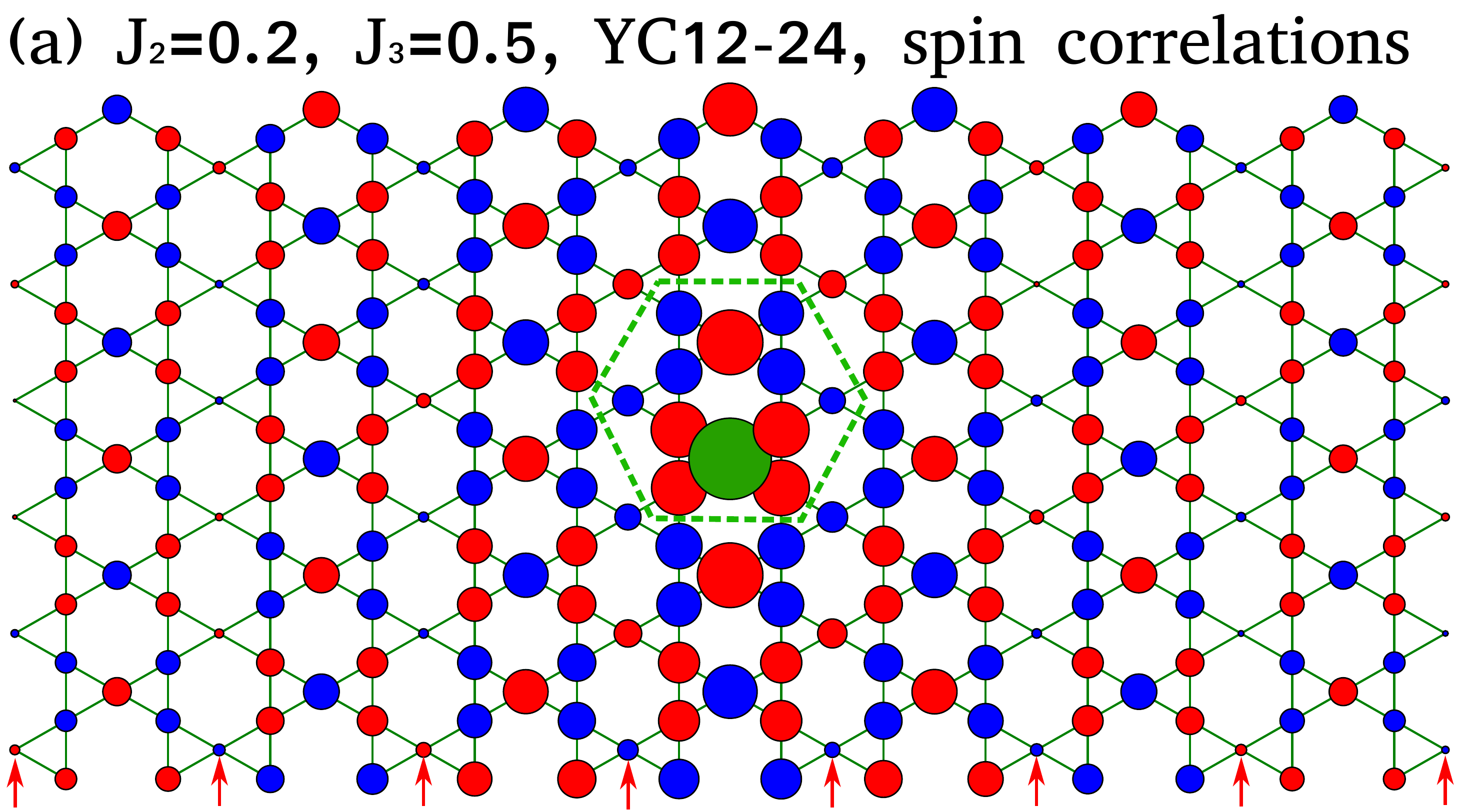}
\includegraphics[width=0.8\linewidth]{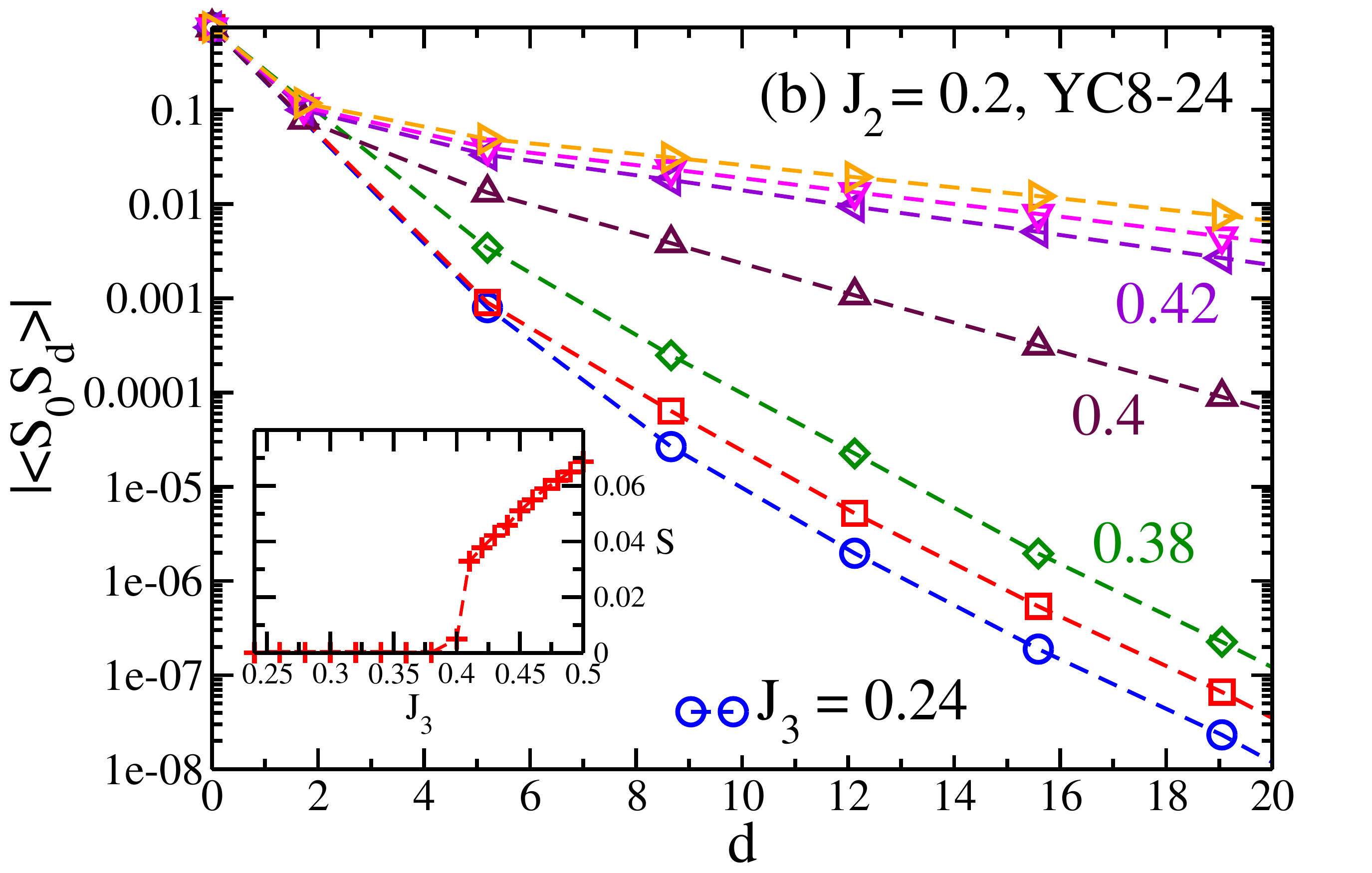}
\includegraphics[width=0.8\linewidth]{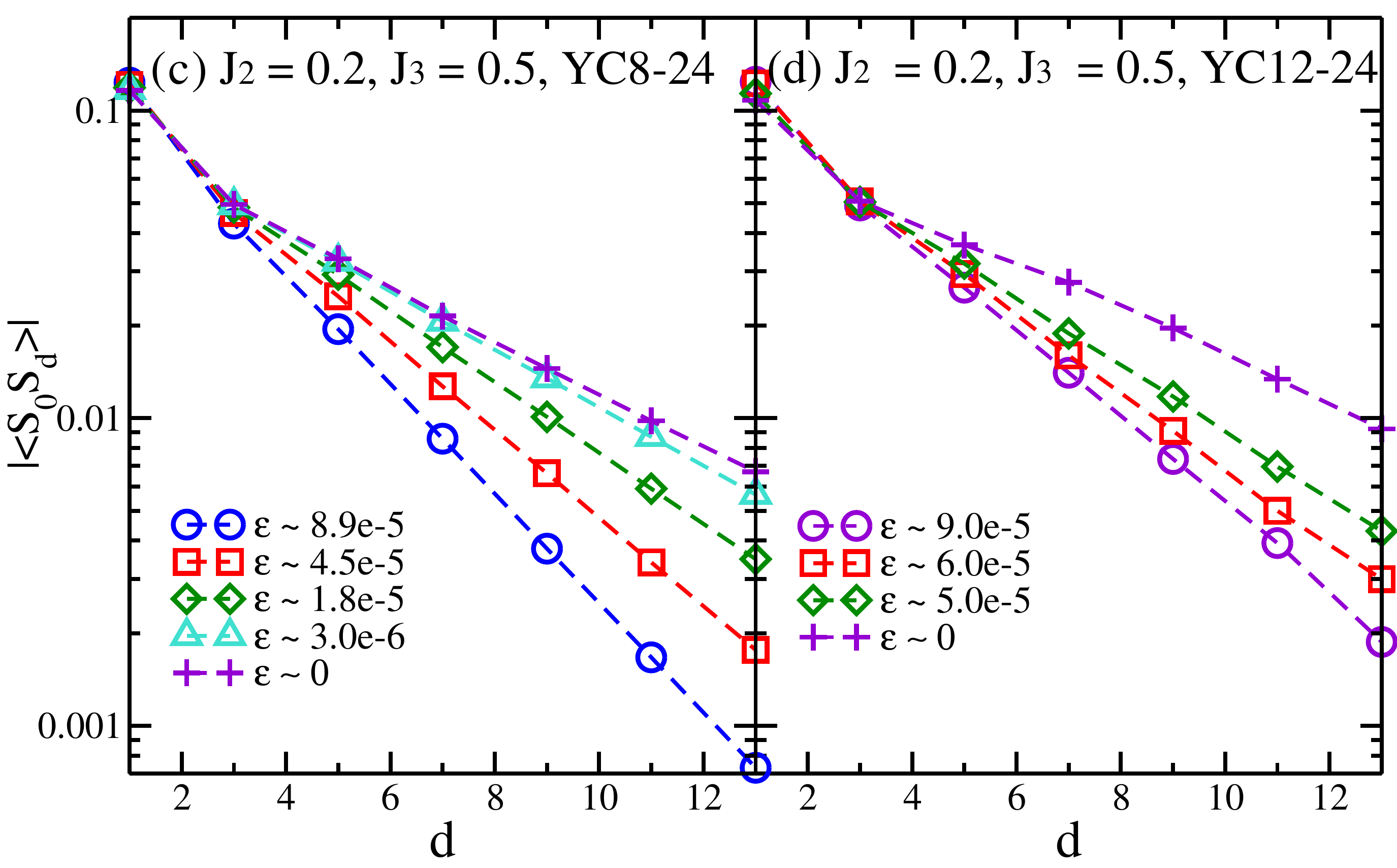}
\includegraphics[width=0.8\linewidth]{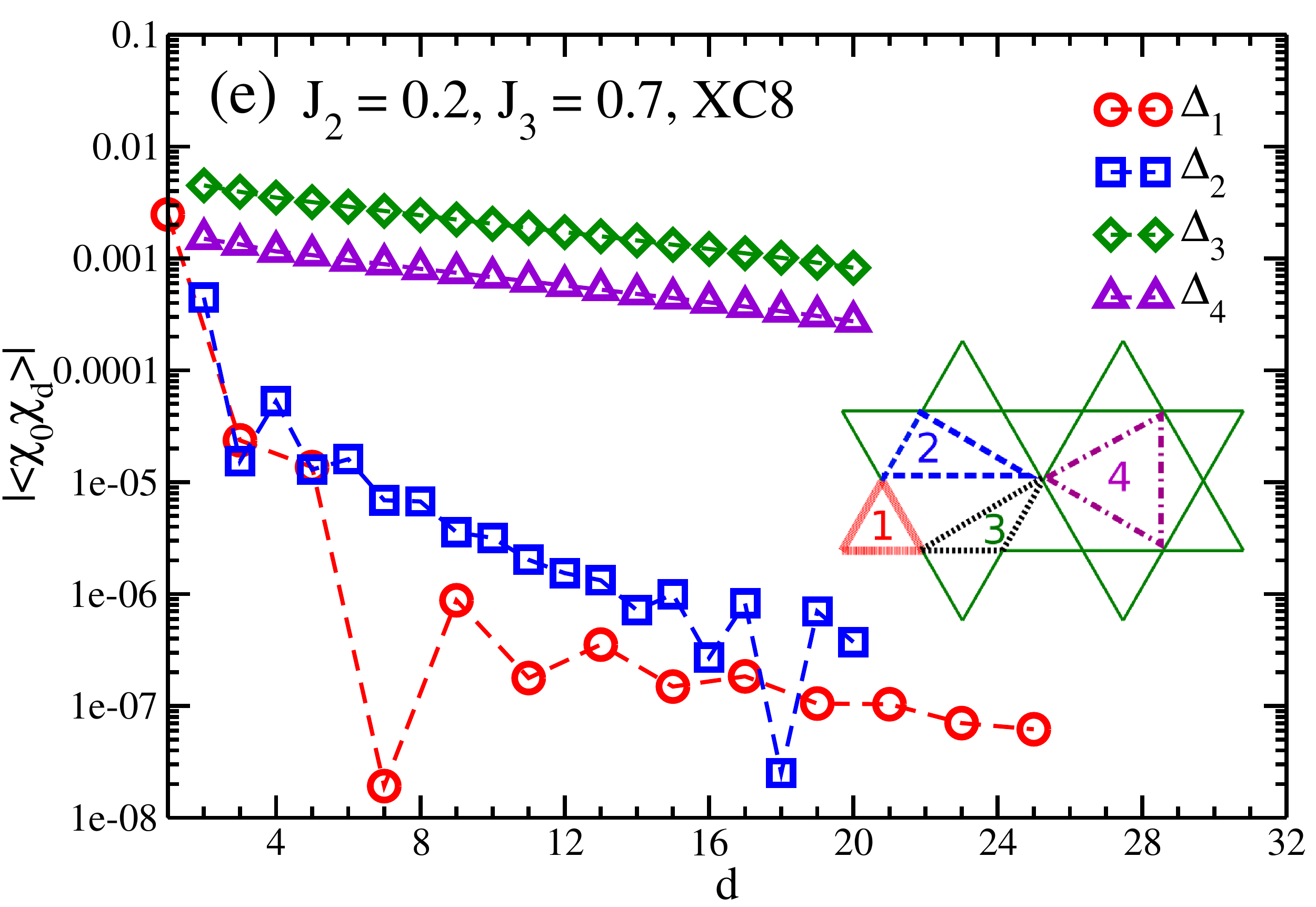}
\caption{Numerical studies of the \textit{cuboc1} phase.  All plots in
  this figure use $J_2=0.2$.  In (a) we show the spin correlations for
  a central region of the YC12-24 cylinder with $J_3 = 0.5$, following
  the same conventions as Fig.~\ref{pattern}.  The dashed hexagon
  indicates the 12-site unit cell. Panel (b) shows log-linear plots of
  the spin correlations for various $J_3$ values on the YC8-24
  cylinder (here $J_3 = 0.24, 0.3, 0.38, 0.4, 0.42, 0.45, 0.5$ for the
  successive curves with increasing values of the correlations).  The
  inset plots the $J_3$ dependence of the long-distance spin
  correlation $S$. Plots (c) and (d) compare the spin correlation for
  $J_3=0.5$ on the YC8-24 and YC12-24 cylinders with different
  truncation errors $\varepsilon$.  The data with plus symbol give the
  results of an extrapolation to zero truncation error.  Panel (e)
  contrasts the correlations of the chirality on the four different
  types of triangles (as shown in the inset) well into the
  \textit{cuboc1} phase for $J_3 = 0.7$ on the XC8-24 cylinder.  The
  correlations on the $\Delta_1, \Delta_2$ triangles are very small
  and sometimes change sign.  This is probably consistent with zero
  spontaneous chirality on these triangles in the thermodynamic
  limit.}\label{cuboc1}
\end{figure}

We first however verify the magnetic order of the \textit{cuboc1}
state by studying the spin-spin correlation function.  The spin
correlation pattern is, in sign and magnitude, consistent with the
\textit{cuboc1} state for several cylindrical geometries, provided
they are chosen compatible with the enlarged unit cell of this
state. An example is shown in Fig.~\ref{cuboc1}(a), where a
\textit{cuboc1} pattern with a 12-site unit cell indicated by the
dashed hexagon is clearly seen.  A characteristic feature is that the
spin correlations in the columns denoted by the red arrows are small
and decay quickly.  This follows naturally from the classical picture
of the \textit{cuboc1} state, because these spins are perpendicular to
the reference spin.

To quantify the magnetic ordering, we study the evolution of the spin
correlations with increasing $J_3$.  An example is shown in
Fig. \ref{cuboc1}(b) for $J_2 = 0.2$, $0.24 \leq J_3 \leq 0.5$ on the
YC8-24 cylinder.  One sees that the spin correlations decay quite fast
for $J_3 < 0.4$, consistent with the gapped CSL shown in
Fig.~\ref{chiral}(b). At $J_3 = 0.4$, the spin correlations are
sharply enhanced and approach finite values at long distance for
$J_3 > 0.4$.  We define $S$ as the square root of the long-distance
spin correlations in Fig. \ref{cuboc1}(b)
$S \equiv \sqrt{|\langle S_0 \cdot S_d \rangle|}$ ($d$ is the longest
distance), and plot it versus $J_3$ in the inset, which shows a jump
of $S$ from zero to a finite value at $J_3 \simeq 0.4$.  The abrupt
simultaneous onset of spin order and vanishing chiral order on small
triangles (Fig.~\ref{chiral}(b)), together indicate a phase transition
from the CSL to a magnetically ordered phase.

To be fully confident of magnetic ordering, we must consider finite
size effects. We compare the spin correlations on YC8 and YC12
cylinders (the XC12 cylinder is incompatible with \textit{cuboc1}
state).  Interestingly, the \textit{cuboc1} state has significantly
enhanced entanglement entropy -- a point which we return to below --
which prevents us from obtaining fully converged results on YC12
cylinder.  Therefore, we instead compare the spin correlations on the
YC8 and YC12 cylinders as obtained with similar truncation errors. As
shown in Figs. \ref{cuboc1}(c) and \ref{cuboc1}(d) for
$J_2 = 0.2, J_3 = 0.5$, the spin correlations grow with decreasing
truncation error (increasing $M$) in both systems.  For similar
truncation errors, the correlation length $\xi$ on the wider YC12
cylinder is always larger than that on the YC8 cylinder.
Consequently, the converged spin correlations (obtained from
extrapolation with respect to truncation error, as shown by the plus
symbol) on the YC12 cylinder are stronger than those on the YC8
cylinder.  The growing spin correlation length is consistent with the
presence of magnetic order in two-dimensional thermodynamic limit.

Now we justify the claim that the CSL {\em cannot} be regarded as a
quantum fluctuating \textit{cuboc1} state.  To see this, we first
consider the pattern of scalar spin chirality in the \textit{cuboc1}
state.  Classically, the spins in triangles $\Delta_1$ and $\Delta_2$
are {\em coplanar}, so these possess zero scalar spin chirality.  Spin
chirality is instead concentrated in triangles $\Delta_3$ and
$\Delta_4$, where the spins are non-coplanar \cite{PRL_108_207204}.
We indeed see precisely this behavior in the numerical calculations of
chirality correlations, which are large and consistent with long-range
chiral order {\em only} for triangles $\Delta_3$ and $\Delta_4$, as
shown in Fig.~\ref{cuboc1}(e).  This is why the chirality calculated
for the small ($\Delta_1$) triangles in Fig.~\ref{chiral}(b) jumps to
zero in the \textit{cuboc1} state.  In Fig.~\ref{cuboc1}(e), we see
some very small residual chirality correlations on the type 1 and 2
triangles, but these are consistent with short-range correlations,
which are always non-zero and do not indicate symmetry breaking. The
absence of scalar spin chirality in the small triangles in the
\textit{cuboc1} state reflects an invariance of this state under a
combined $C_2$ rotation in spin space (about an axis through two
antipodal points on the cuboctahedron) and a real space reflection
through a plane bisecting a column of small triangles.  The CSL state
breaks this symmetry.  Hence the two phases are {\em symmetry
  distinct} even beyond the presence of spin ordering.

\begin{figure}
\includegraphics[width=1.0\linewidth]{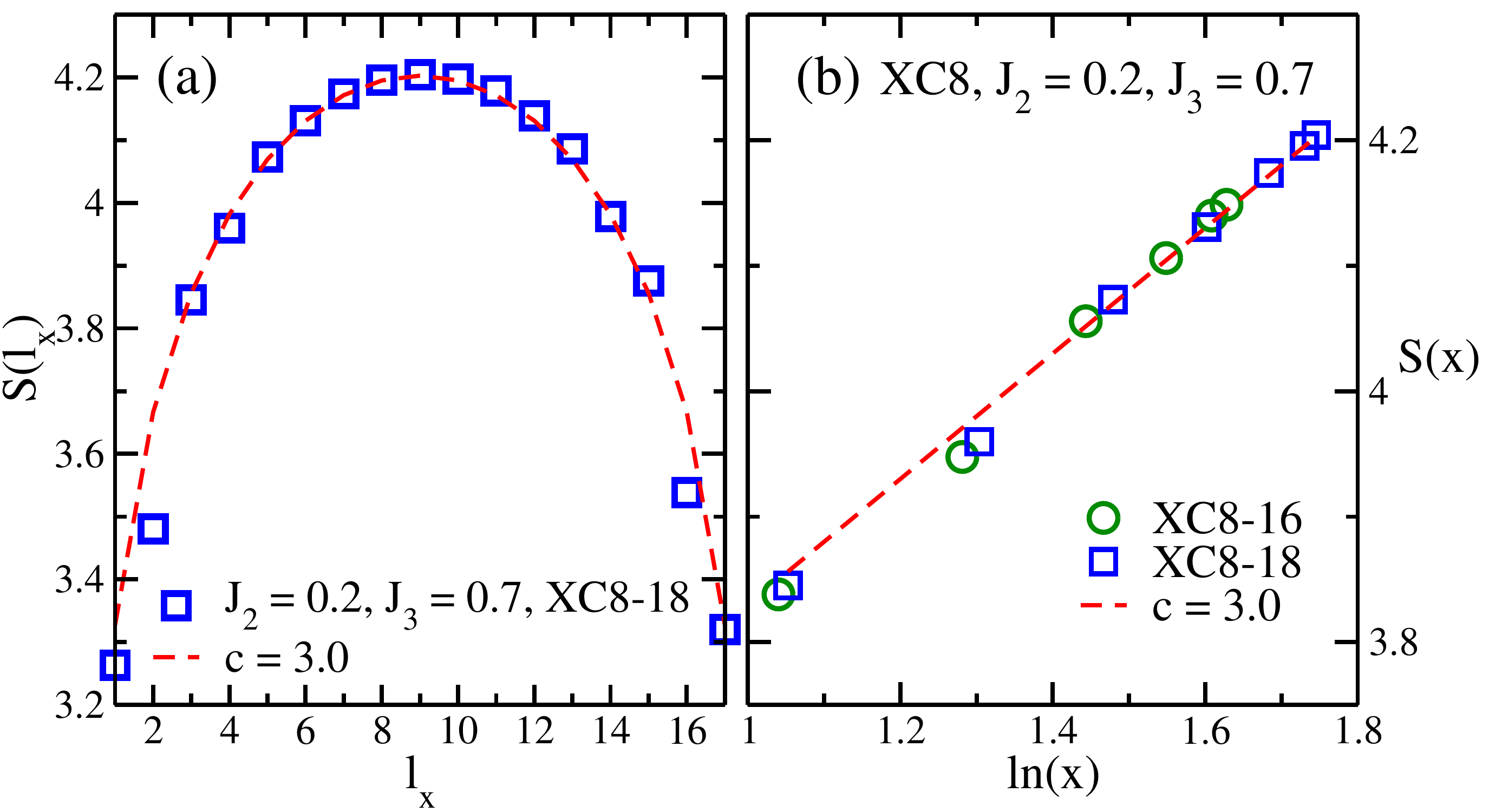}
\caption{Entanglement entropy in the \textit{cuboc1} phase.  In (a)
  the entanglement entropy versus the sub-system length $l_x$ for
  $J_2=0.2, J_3=0.7$ on the XC8-18 cylinder is fitted using the
  conformal field theory formula
  $S(l_x) = (c/6)\ln[(L_x/\pi)\sin(l_x\pi/L_x))] + g$ with
  $c=3.0, g=3.3$. In (b), the entanglement entropy is plotted versus
  $\ln x$ ($x = (L_x/\pi)\sin(l_x\pi/L_x)$) for $J_2=0.2, J_3=0.7$ on
  XC8-16 and XC8-18 cylinders.  The dashed line is the fit curve used
  shown in (a).}\label{cc}
\end{figure}

Finally, we return to the entanglement entropy in the \textit{cuboc1}
state.  The large entropy may be understood from general arguments.
In the two-dimensional limit, the \textit{cuboc1} phase fully breaks
SU(2) spin symmetry, and so has {\em three} gapless Goldstone modes (a
number equals to the number of generators of SU(2)).  This is described
field-theoretically by an 2+1-dimensional SO(3) matrix non-linear sigma model \cite{sigma}.  
If we now imagine placing the \textit{cuboc1} state on a (compatible)
cylinder, the momentum along the circumferential direction $k_y$ becomes
quantized, and we na\"ively expect three gapless one-dimensional
bosonic modes with $k_y=0$.  In general, these modes are interacting,
and for long cylinders fluctuate strongly and are expected to open up
a gap, since the non-linear sigma model in 1+1-dimensions is
asymptotically free.  However, this gap is exponentially small when
the cylinder circumference is large, and so we can expect a wide
regime in which the cylinder behaves like a system of {\em three} gapless
free bosonic modes. 

For a general free gapless bosonic system (actually any conformal
field theory) in 1+1-dimensions, the entanglement entropy of a bipartition into two
halves follows the area law
$S(l_x) = (c/6)\ln[(L_x/\pi)\sin(l_x\pi/L_x))] + g$ \cite{central}, where $c$ is the
characteristic central charge of system, $g$ is a nonuniversal
constant reflecting short-range entanglement, and $l_x$ and $L_x$ are
the length of subsystem and the whole system, respectively. The
non-linear sigma model argument above implies $c=3$.  Thus the large
entanglement entropy of the \textit{cuboc1} state on cylinders could
be attributed to its Goldstone mode structure.

We verify this numerically in more detail, and find behavior
consistent with this prediction. An example of the $l_x$ dependence of
entropy is shown in Fig.~\ref{cc}(a) for $J_2=0.2,J_3=0.7$ on the
XC8-18 cylinder. We bipartition the system column by column, and
denote the number of columns as $l_x$.  The entropy fits quite well
using the area law behavior with $c=3.0, g=3.3$.  In Fig.~\ref{cc}(b),
we plot the same data versus $\ln[(L_x/\pi)\sin(l_x\pi/L_x))]$, where
the slope of the dashed line determines the central charge. We find
that the entropy on XC8-16 cylinder also follows the same central
charge $c=3.0$.

\section{Quantum phase transitions}

It is interesting to study the phase transitions between the
well established topological CSL and other phases 
surrounding it. Continuous phase transitions from such a
topologically ordered phase are of general interest as examples of
unconventional quantum criticality \cite{Science_303_1490, PRB_70_144407}.  
Thus we attempt to establish if any of the transitions in our 
system are indeed continuous.

\begin{figure}
\includegraphics[width=1.0\linewidth]{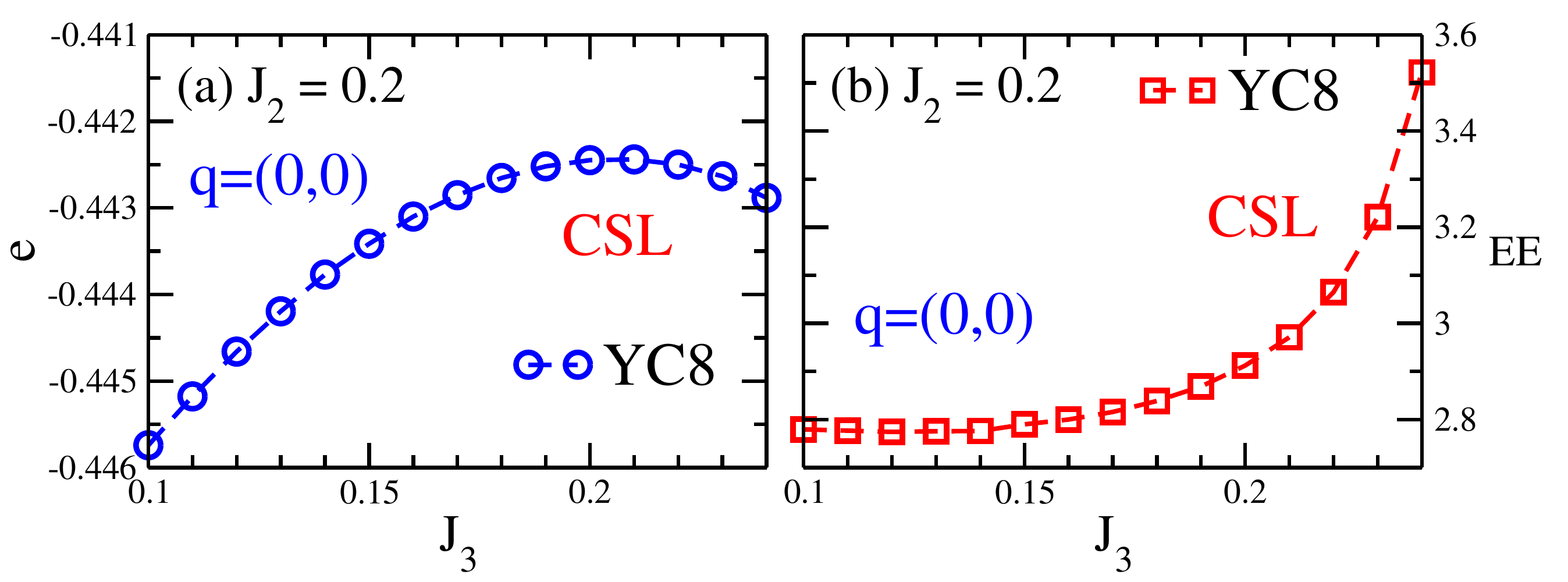}
\includegraphics[width=1.0\linewidth]{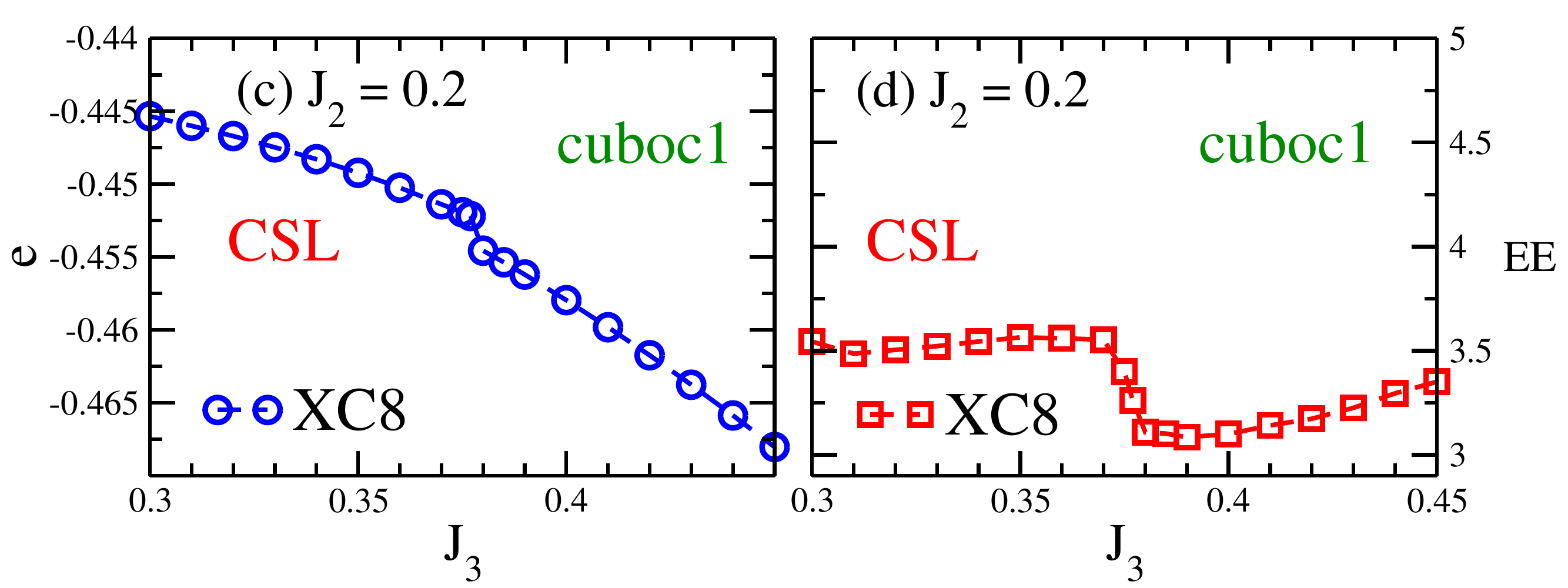}
\includegraphics[width=1.0\linewidth]{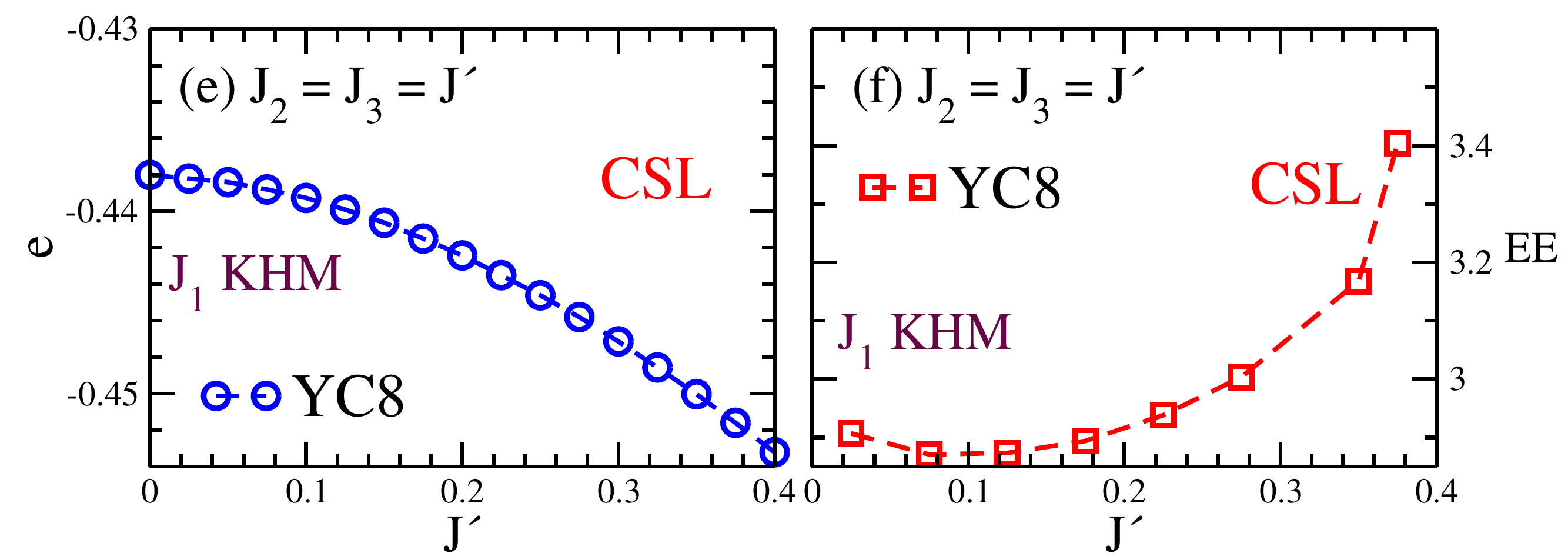}
\caption{Scans of ground state energy and entanglement entropy of a
  bipartition into equal halves, as probes of quantum phase
  transitions.  Panels (a) and (b) show the $J_3$ dependence for
  $J_2 = 0.2$ on the YC8 cylinder over a region spanning the
  transition from the CSL to the $q=(0,0)$ N\'{e}el phase.  No sharp
  features are observed, suggesting a continuous phase transition.  In
  panels (c) and (d) the same quantities are shown for for $J_2 = 0.2$
  on the XC8 cylinder, to probe the transition from the CSL to the
  \textit{cuboc1} phase.  In this case an abrupt feature is observed,
  suggesting a first order transition.  In the final two panels, (e)
  and (f), the system is scanned with $J_2=J_3=J^{\prime}$ on the YC8
  cylinder, to study the transition from the time-reversal invariant
  QSL to the CSL.  In this case, the transition again appears
  continuous.  }\label{transition}
\end{figure}

First, we consider the phase transition from the CSL to the $q=(0,0)$
N\'{e}el phase.  For $J_2 = 0.2$ on YC8 cylinder, we find the
transition occurs at about $J_3 \simeq 0.2$, based upon the behavior
of chiral and spin correlations in Fig.~\ref{chiral}(b) and
Fig.~\ref{spin}.  To gauge the order of the transition, we plot in
Figs.~\ref{transition}(a) and \ref{transition}(b) the $J_3$ dependence
of the ground-state energy and entanglement entropy for $J_2 = 0.2$ in
a range spanning the $q=(0,0)$ to CSL transition on the YC8-24
cylinder.  We find that both the ground-state energy and entropy vary
smoothly with $J_3$, indeed so smoothly that a transition cannot be
identified from these data.   This suggests the CSL to N\'eel transition
may be continuous.  However, we should caution that the absence of
sharp features is not evidence for criticality -- which in any case
would be difficult to verify on the small systems studied here.  It
does indicate that the CSL to N\'eel transition is not strongly first order.

This is in contrast to the transition from the CSL to the
\textit{cuboc1} phase.  As shown in Figs.~\ref{transition}(c) and
\ref{transition}(d) of the results on XC8 cylinder for $J_2 = 0.2$, we
find both the energy and the entropy have a sharp change at $J_3
\simeq 0.38$, which are also observed on YC8 cylinder at $J_3 \simeq
0.4$.  Note that for a large system, the theoretical expectation at a
first order transition between these two phases is a slope
discontinuity in the ground state energy and a jump in the
entanglement entropy, both of which are compatible with
Figs.~\ref{transition}(c) and \ref{transition}(d).  The sharp changes observed in these
quantities are also consistent with the sudden drop of chiral
correlation in Fig.~\ref{chiral}(b) as well as the enhancement of spin
correlations in Fig.~\ref{cuboc1}(b).  All these results indicate a strong
first-order transition from the CSL to the \textit{cuboc1} phase.
The first order nature of this transition is another indication that
the CSL phase should not be regarded as a quantum fluctuating
descendent of the \textit{cuboc1} phase, as discussed above in
Sec.~\ref{sec:textitcuboc1-phase}.  

Next we consider the $J_1$-$J^{\prime}$ model with $J_2=J_3=J'$ to
investigate the transition from the CSL to QSL ground state of the
pure nearest-neighbor KHM.  The latter phase itself is under debate,
and may be a gapped $Z_2$ QSL \cite{Science_332_1173} or a gapless
QSL \cite{XY_model} of either $Z_2$ or $U(1)$ type\cite{PRL_98_117205, PRB_87_060405}.
By studing the chiral correlations, the transition point is determined to be $J^{\prime}
\simeq 0.07$ as shown in Fig.~\ref{phase}(a). In Figs.~\ref{transition}(e) and \ref{transition}(f),
we find that both the energy and entropy appear smooth as a function of
$J^{\prime}$, leaving open the possibility of a continuous transition
of the nearest-neighbor QSL state into the CSL phase \cite{1407.2740}.

\section{Summary and Discussion}

We have studied the competing quantum phases of the spin-$1/2$
$J_1$-$J_2$-$J_3$ KHM for $0 \leq J_2 \leq 0.25$ and
$0.0 \leq J_3 \leq 1.0$ by DMRG simulations.  As
shown in Fig.~\ref{phase}(a), we find five phases: a
time-reversal symmetric QSL state continuously connected to that of the nearest-neighbor
Heisenberg model, a $q=(0,0)$ N\'{e}el phase, a chiral spin liquid
(CSL) phase, a non-coplanar \textit{cuboc1} phase, and a VBC phase.
The CSL phase seems to arise as a result of quantum fluctuations
around the line of classical degeneracy between the two types of
classical order: the $q=(0,0)$ N\'{e}el phase and \textit{cuboc1}
phase.   The chirality structure of the CSL and \textit{cuboc1} phases
are distinctly different, and indeed we find a strong first order
phase transition between them.  

Both the quantum phase transition between the CSL and the $q=(0,0)$
N\'eel state, and that between the CSL and the time-reversal symmetric
QSL, are quite smooth and consistent with continuous behavior.  If
continuous, these could be interesting examples of unconventional
quantum critical points \cite{Science_303_1490, PRB_70_144407}.  It is
not clear even what to expect for the universal field theories for
these phase transitions from the theory of QSLs.  The nature of the
time-reversal symmetric QSL itself is controversial, making it hard to
speak definitively about that transition.  If we suppose that the QSL
itself is of the gapless U(1) Dirac type \cite{PRL_98_117205}, then
this transition could be understood as simple ``chiral symmetry
breaking''-type transition in which a scalar mass gap appears for the
Dirac fermions \cite{PRB_39_11413,hsf}. The mechanism for generation of
an appropriate Chern-Simons term to describe the universal aspects of
the CSL from such a Dirac mass is
well-known \cite{PRB_39_11413,Redlich}. However, it is not clear that
the proposed U(1) Dirac state is even stable as a phase.  At this
point, we have only some speculative ideas for the field theories
that might describe transitions from a $Z_2$ version of the
time-reversal symmetric QSL liquid state, or from the $q=(0,0)$ N\'eel
state, to the CSL.  We suggest this may be a possibly fruitful problem
for future research.


We acknowledge the stimulating discussions with L.~Messio, W.~J.~Hu, T.~Grover,
Z.~Y.~Weng, and X.~G.~Wen.  This research is supported by the National
Science Foundation through grants DMR-1408560 (S.S.G.), DMR-1206809
(L.B.), and the U.S. Department of Energy,
Office of Basic Energy Sciences under grant No. DE-FG02-06ER46305
(D.N.S,W.Z.).
\\

\textit{Note added.}---Upon finalizing the manuscript we noticed a
recent preprint reporting a DMRG \cite{1410.7911} study of the
$J_1$-$J_2$  Heisenberg model, and also finds that the $q=(0,0)$
order emerges for $J_2 > 0.2$.  We also noticed a preprint on
variational Monte Carlo studies \cite{1410.7359} of the same model
based on the Gutzwiller projected fermion wavefunction, which claims
that the $U(1)$ Dirac spin liquid may be stable in a region with
finite $J_2 > 0$.



\begin{thebibliography}{99}
\bibitem{convergence} While the calculations for $L = 2, 4$ are easy to converge, we find that $m^2$ continues to grow with
increasing kept states for $L = 6$. By keeping the states up to about $24000$, the increase of $m^2$
is negligible and the results are close to convergence.
\end{thebibliography}
\end{document}